\documentclass[%
 reprint,
 amsmath,amssymb,
 aps,
 pra,showkeys,
]{revtex4-2}

\usepackage{graphicx}% Include figure files
\usepackage{dcolumn}% Align table columns on decimal point
\usepackage{braket}
\usepackage{bm}% bold math
\usepackage{bbm}% bold math
\usepackage{array}
\usepackage{subcaption}
\usepackage[hidelinks,breaklinks]{hyperref}
\begin{document}

\preprint{APS/2309.10509}

\title{Quantum-inspired Tensor Network for QUBO, QUDO and Tensor QUDO Problems with $k$-neighbors}% Force line breaks with \\

\author{Sergio Muñiz Subiñas}
 \email{sergio.muniz@itcl.es}
\affiliation{Instituto Tecnológico de Castilla y León, Burgos, Spain}
% \orcidlink{0009-0008-7590-0149}

\author{Alejandro Mata Ali}
 \email{alejandro.mata@itcl.es}
\affiliation{Instituto Tecnológico de Castilla y León, Burgos, Spain}
% \orcidlink{0009-0006-7289-8827}

\author{Jorge Martínez Martín}
 \email{jorge.martinez@itcl.es}
\affiliation{Instituto Tecnológico de Castilla y León, Burgos, Spain}
% \orcidlink{0009-0004-9336-3165}

\author{Miguel Franco Hernando}
 \email{miguel.franco@itcl.es}
\affiliation{Instituto Tecnológico de Castilla y León, Burgos, Spain}
% \orcidlink{0009-0009-9059-6832}

\author{Javier Sedano}
 \email{javier.sedano@itcl.es}
\affiliation{Instituto Tecnológico de Castilla y León, Burgos, Spain}
% \orcidlink{0000-0002-4191-8438}

\author{Ángel Miguel García-Vico}
 \email{agvico@ujaen.es}
\affiliation{Andalusian Research Institute in Data Science and Computational Intelligence (DaSCI), University of Jaén, 23071 Jaén, Spain}
% \orcidlink{0000-0003-1583-2128}

\date{\today}% It is always \today, today,
             %  but any date may be explicitly specified

\begin{abstract}
This work presents a novel tensor network algorithm for solving Quadratic Unconstrained Binary Optimization (QUBO) problems, Quadratic Unconstrained Discrete Optimization (QUDO) problems, and Tensor Quadratic Unconstrained Discrete Optimization (T-QUDO) problems. The algorithm proposed is based on the MeLoCoToN methodology, which solves combinatorial optimization problems by employing superposition, imaginary time evolution, and projective measurements. Additionally, two different approaches are presented to solve the QUBO and QUDO problems with $k$-neighbors interactions in a lineal chain, one based on 4-order tensor contraction and the other based on matrix-vector multiplication, including sparse computation and a new technique called ``Waterfall''. Furthermore, the performance of both implementations is compared with a quadratic optimization solver to demonstrate the performance of the method, showing advantages in several instances of problems.
\end{abstract}

%Use showkeys class option if keyword
                              %display desired
\maketitle

% \tableofcontents
\section{Introduction}
Combinatorial optimization is a branch of mathematics that plays a central role in a wide range of real-world problems, including scheduling~\cite{flex_jssp}, logistics~\cite{VRP}, and resource allocation~\cite{Assignment}. Many of these problems are classified as NP-hard problems~\cite{combinatorial_optimization_theory_papa,combinatorial_optimization_theory_garey}, which implies that exact methods are computationally infeasible for large-scale instances. Consequently, the development of algorithms capable of producing near-optimal solutions constitutes an important objective of contemporary research.

An interesting family of problems is the class of quadratic optimization problems~\cite{quadratic}, in which the decision variables are restricted to take values from a finite set of discrete numbers. Many important combinatorial problems, such as the Traveling Salesman Problem (TSP) or the Knapsack Problem, can be formulated in this way~\cite{gonzálezbermejo2022gpsnewtspformulation, knapsack_qubo}, which underscores the practical relevance of this class of problems. Within the family of quadratic integer optimization problems, several variants can be distinguished. When variables are restricted to binary values, the problem is known as Quadratic Unconstrained Binary Optimization (QUBO)~\cite{quadratic_qubo}. If variables can take arbitrary integer values from a finite set, the formulation is known as Quadratic Unconstrained Discrete Optimization (QUDO) or Quadratic Integer Optimization (QIO).

Current state-of-the-art methods for solving these problems often rely on different classical optimization techniques. Exact approaches, including branch-and-bound~\cite{brach_and_bound} and linear programming~\cite{Linear}, can obtain optimal solutions for small and medium-sized instances, but their scalability is limited. For larger problems, it is common to employ heuristic and metaheuristic strategies, such as simulated annealing~\cite{simmulated_annealing_kirk}, tabu search~\cite{tabu_search_glover} or genetic algorithms \cite{genetic_holland}. More recently, hybrid techniques that integrate local search with mathematical programming have shown strong performance. Some of the most used and specialized solvers are CPLEX~\cite{cplex2024}, GUROBI~\cite{gurobi2024} or OR-Tools~\cite{ortools2024}. However, some of them have high associated costs and hard integration due to licensing, require powerful machines for parallelization, or are less adapted to certain types of hard and specific problems.

The method employed in this work is based on tensor networks \cite{tensornetworksnutshell,lecturesquantumtensornetworks,Orus_TN}, a widely developed mathematical formalism in condensed matter physics~\cite{Orus_Many_body}. Tensor networks are particularly powerful because they enable the efficient representation and manipulation of high‑dimensional data structures. Their strength lies in the ability to exploit the locality and constrained connectivity patterns inherent in the problem. In this study, the adopted tensor network technique is Modulated Logical Combinatorial Tensor Networks (MeLoCoToN)~\cite{melocoton}, a methodology originally inspired by quantum computation concepts such as superposition and imaginary time evolution, but implemented entirely in a classical computational setting.

Building on these considerations, this paper introduces a tensor network capable of exactly formulating general QUBO and QUDO problems. It provides the exact equation that solves the k-nearest‑neighbor case, developed in two alternative forms: one employing higher‑order tensor contractions and another based on matrix–vector multiplications. In addition, it presents a new technique that improves the process of determining the optimal assignment of variables within the MeLoCoToN algorithm, reducing the computational memory and the overflow under some circumstances.
Briefly, the main contributions of this work are:
\begin{enumerate}
    \item An exact and explicit equation which solves general QUBO, QUDO and Tensor-QUDO (T-QUDO) problems, based on a tensor network lattice, and two equations for the particular case of the $k$-neighbors interaction in a lineal chain.
    \item Two contraction schemes and their computational complexity for both problems, with a comparative study of which one is better in which problem sizes.
    \item A novel algorithm for the determination of the optimal variable values for MeLoCoToN-type algorithms without contracting all the tensor network and avoiding in some circumstances the overflow due to the imaginary time evolution.
\end{enumerate}
Additionally, it is provided an open-source implementation of both algorithms and the experimentation of the paper in the GitHub repository \href{https://github.com/SergioITCL/QUDO-tensor-network-solver}{https://github.com/SergioITCL/QUDO-tensor-network-solver}.

The remainder of this paper is organized as follows: Sec.~\ref{sec:background} presents the necessary background and formal definitions of QUBO, QUDO, and T-QUDO, and also briefly explains the bra-ket notation and the MeLoCoToN methodology. Sec.~\ref{sec:tn_equation} describes the tensor network formulation used in this work for QUBO, QUDO and T-QUDO problems, the calculation algorithm and their computational complexity. Sec.~\ref{sec:waterfall} introduces the new variation of MeLoCoToN, the waterfall method. Sec.~\ref{sec:experiments} reports the experimental evaluation and comparisons with a baseline solver. Finally, Sec.~\ref{sec:conclusions} concludes the paper with perspectives for future research.

\section{Problem Formulation, Preliminaries and Background}\label{sec:background}
To provide the necessary context, this section reviews the formulations of QUBO, QUDO, and Tensor-QUDO problems. A brief overview of bra–ket notation, along with a summary of the MeLoCoToN methodology, explaining the fundamental principles of the algorithm and its application to combinatorial optimization problems.

\subsection{QUBO, QUDO and Tensor-QUDO formulations}
The QUBO problem consists in finding the binary vector $\vec{x} \in \{0,1\}^n$ that minimizes a cost function of the form
\begin{equation}
C(\vec{x}) = \sum_{\substack{i,j=0 \\ i \leq j}}^{n-1}Q_{ij}x_ix_j.
\end{equation}
The definition of the problem is fully characterized by a symmetric matrix $Q$. An interesting property satisfied by QUBO problems is $x_i^2=x_i$, resulting in the absorption of the linear term into the quadratic component.

The QUDO problem generalizes QUBO by allowing each variable to take integer values from a finite set $x_i \in \{0,1,...,d-1\}$. Unlike QUBO, QUDO does not satisfy the identity $x_i^2=x_i$, which prevents the absorption of the linear term into the quadratic term. The cost function is
\begin{equation}
C(\vec{x}) = \sum_{\substack{i,j=0 \\ i \leq j}}^{n-1}Q_{ij}x_ix_j + \sum_{i=0}^{n-1}D_ix_i,
\end{equation}
being $Q$ a symmetric matrix and $\vec{D}$ a vector.

A more complex extension is the T-QUDO problem, where the cost is the sum of tensor elements selected by the values of the variables. The expression is
\begin{equation}\label{eq:tqudo}
C(\vec{x}) = \sum_{\substack{i,j=0 \\ i \leq j}}^{n-1}\hat{Q}_{i,j,x_i,x_j},
\end{equation}
where $\hat{Q}$ is the cost tensor. The QUBO and QUDO formulations are particular cases where $\hat{Q}_{i,j,x_i,x_j} = Q_{i,j}x_ix_j$. In previous work, similar problems with interactions in a linear chain have been solved~\cite{QUBO_tridiaonal}, in which the matrix $Q$ is tri-diagonal and the cost tensor $\hat{Q}$ has only terms $\hat{Q}_{i,i,j,k}$ and $\hat{Q}_{i,i+1,j,k}$.

Several methods are commonly applied to these types of problems. Exact methods include branch‑and‑bound~\cite{brach_and_bound} and branch‑and‑cut~\cite{branch_cut}, which systematically explore the solution space while pruning suboptimal branches, and cutting‑plane techniques that iteratively tighten the feasible region. Those are the core approaches of some solvers such as CPLEX~\cite{cplex2024} or Gurobi~\cite{gurobi2024}. The most common heuristic techniques are: simulated annealing~\cite{simmulated_annealing_kirk}, which simulates the physical process of physical annealing to escape local minima by accepting worse solutions with a certain probability; tabu search~\cite{tabu_search_glover}, which guides a local search by checking immediate neighbors; and genetic algorithms~\cite{genetic_holland}, which evolve a population of solutions through selection, crossover, and mutation operators \cite{heuristic_comparation}. In addition, some emerging technologies propose novel algorithms that show promising computational advantages over traditional methods. For example, quantum computing~\cite{nielsen, glover2019, forecastingquantumcomputing} introduces algorithms that in theory can solve certain problems more efficiently. Quantum annealing \cite{quantum_annealing_kadowaki}, an algorithm based on the adiabatic theorem that performs the evolution of the quantum state under some conditions in order to obtain the ground state of the Ising Hamiltonian related to the QUBO problem. Or the Quantum Approximate Optimization Algorithm (QAOA)~\cite{QAOA} which discretizes the adiabatic theorem and performs an optimization process to obtain the ground state as in the quantum annealing.

\subsection{Bra-ket notation}
In the tensor network formalism, bra-ket notation is a concise way of representing tensors and their manipulation. In this work, only a brief overview of the notation is needed; for a more precise and detailed treatment, the reader should consult the references listed in the bibliography \cite{cohen}.

A ket $\ket{\psi}$ denotes a column vector in a vector space. A bra $\bra{\psi}$ denotes the hermitian conjugate of the ket. The inner product between two vectors is denoted as a bra-ket operation $\langle \phi | \psi \rangle$, resulting in a number. The outer product between two vectors is denoted as a ket-bra operation $\ket{\phi} \bra{\psi}$, representing a matrix.

For a system of one variable, let $\mathcal{H}_d$ be a $d$-dimensional vector space, its standard basis consists of $d$ orthonormal vectors $\{\ket{0}, \ket{1},...,\ket{d-1}\}$, where each $\ket{x}$ is represented as a column vector with a 1 in position $x$ and 0 elsewhere. It is possible to express any vector $\ket{\psi}$ as
\begin{equation}
    \ket{\psi} = \sum_{x=0}^{d-1} \alpha_x \ket{x}, \quad \alpha_x \in \mathcal{C}.
\end{equation}
For a system of $n$ variables, the Hilbert space is $\mathcal{H} = \mathcal{H}_{d_0} \otimes \mathcal{H}_{d_1} \otimes \dots \otimes \mathcal{H}_{d_{n-1}}$, where $\otimes$ denotes the tensorial product, and the dimension of the total space is
\begin{equation}
    \dim(\mathcal{H}) = \prod_{i=0}^{n-1} d_i.
\end{equation}
The standard basis in this situation consists of all possible tensor products of the local basis vectors
\begin{equation}
\ket{\vec{x}} = \ket{x_0 x_1 \dots x_{n-1}} = \ket{x_0} \otimes \ket{x_1} \otimes \dots \otimes \ket{x_{n-1}}.
\end{equation}
Each $\ket{x_0 x_1 \dots x_{n-1}}$ can be interpreted as a tensor of $n$ indices, which is very convenient to represent in the tensor network formalism.
It is possible to express an arbitrary tensor within $\mathcal{H}$ using this basis

\begin{equation}
    \begin{gathered}
    \ket{\psi} =
    \sum_{x_0=0}^{d-1}  \dots \sum_{x_{n-1}=0}^{d-1}
    \alpha_{x_0,\dots,x_{n-1}} \;
    \ket{\vec{x}}, \\
    \quad \alpha_{x_0,\dots,x_{n-1}} \in \mathbb{C}.
    \end{gathered}
\end{equation}
    If the state satisfies the condition of being expressible as a tensor product of individual subsystems, it can be written as
\begin{equation}
    \ket{\psi} =
    \bigotimes_{i=0}^{n-1} \left(
    \sum_{x_i=0}^{d-1} \alpha_{i,x_i} \ket{x_i}
    \right),
    \quad \alpha_{i,x_i} \in \mathbb{C}.
\end{equation}
\subsection{MeLoCoToN formalism}
The MeLoCoToN method~\cite{melocoton} is an approach that converts any combinatorial problem, such as optimization, inversion, or constraint satisfaction, into an exact and explicit tensor network equation. The application of this formalism to a combinatorial optimization problem consists of the following steps:
\begin{enumerate}
    \item Creation of a classical logical circuit. This circuit receives as input a possible solution and associates an internal number exponentially lower with its cost.
    \item Tensorization of the logical circuit of the problem, transforming every logical operator into a restricted tensor. This provides the tensor network that represents the tensor $T$ with elements
    \begin{equation}
        T = e^{-\tau C(\vec{x})}\ket{x}\bra{x}.
    \end{equation}
    \item Initialization and half partial trace, to consider at the same time all possible inputs and outputs given a set of values for the variable to determine. This represents the tensor
    \begin{equation}
        \ket{P^{i}} = \sum_{j=0}^{d_i-1}\left(\sum_{\vec{x}|x_i=j}e^{-\tau C(\vec{x})}\right)\ket{j}.
    \end{equation}
    \item Extraction of variable values with an argmax function on the $P^i$ tensor, or with binary extraction with $(-1,1)$ vectors, to obtain a tensor network that contracted is equal to $\Omega_i(\tau)$.
    \item Iteration to determine all the variables.
\end{enumerate}

The final expression provided by this methodology usually is an explicit equation of the type
\begin{equation}
    x_i = \lim_{\tau\rightarrow\infty}H(\Omega_i(\tau)),
\end{equation}
being $H(\cdot)$ the Heaviside step function. This equation is exact in the limit. However, outside the limit can be computed with contraction schemes and provide an approximate solution, good for large enough values of $\tau$.

\section{Tensor network algorithm}\label{sec:tn_equation}
This section provides the main contribution of this work, the tensor network algorithm for solving QUBO, QUDO, and Tensor QUDO problems. Sec.~\ref{ssec:general qudo} focus in the resolution of QUDO problems and Sec.~\ref{ssec: kneigh qudo} presents the resolution for QUDO problems with interaction of $k$-neighbors in a linear chain. Both sections include the deduction of the tensor network equation, the computation algorithm, and its computational complexity. Sec.~\ref{ssec: tensor_qudo} presents the generalization of the previous methods for the Tensor QUDO problems, and Sec.~\ref{ssec: exact explicit} presents the final exact and explicit tensor network equation expression summarized.

\subsection{General QUDO problems}\label{ssec:general qudo}

\subsubsection{Tensor network deduction}\label{sssec:gen deduction}
\begin{figure}
    \centering
    \includegraphics[width=0.95\linewidth]{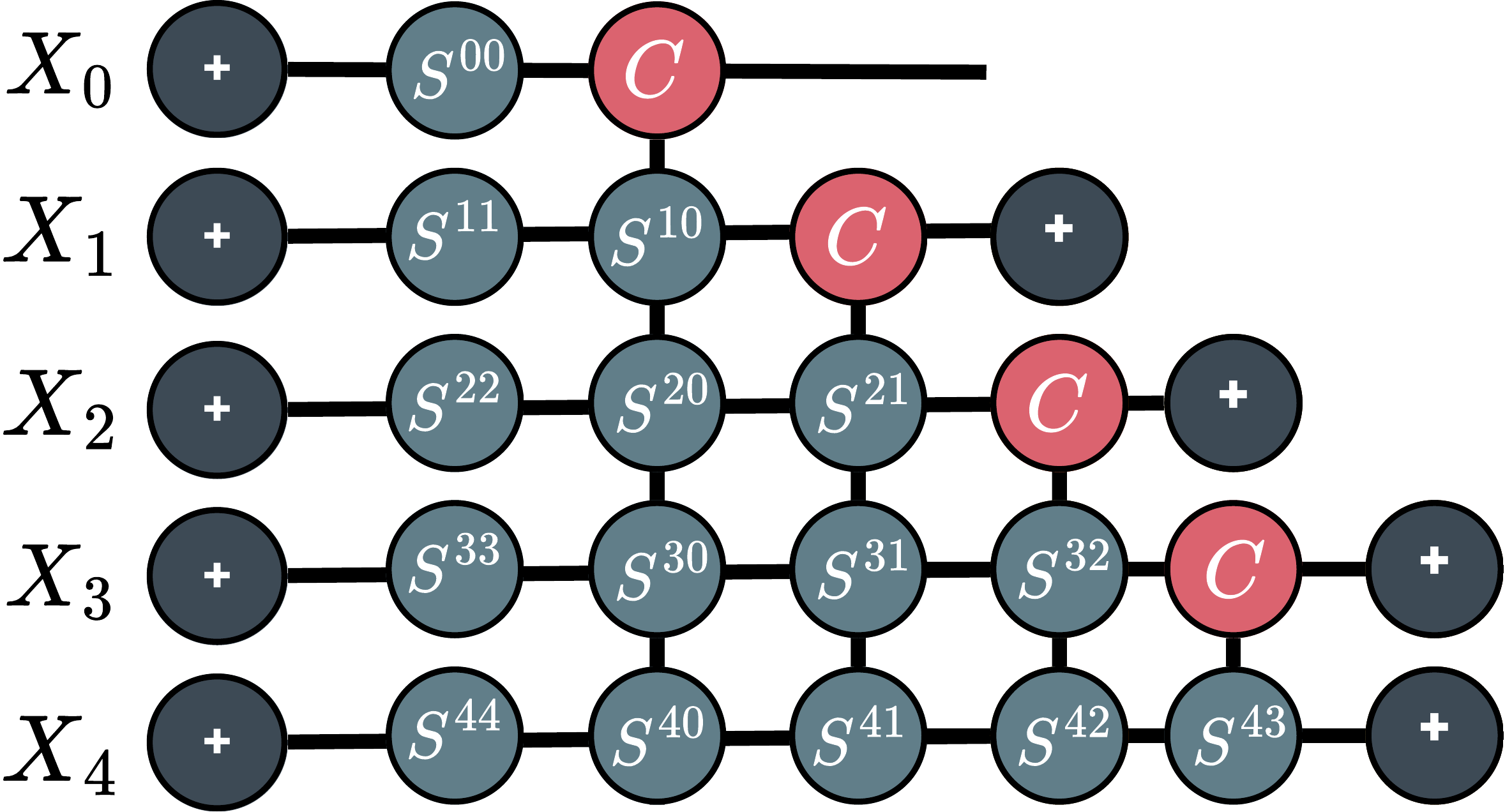}
    \caption{Tensor network for a dense QUDO problem with 5 variables. The mathematical description of each tensor is in Appendix~\ref{appendix:section-1}.}
    \label{fig: qudo_tn}
\end{figure}

As in other algorithms based on MeLoCoToN \cite{melocoton}, the objective is to generate a tensor network that represents the tensor that contains every possible combination with an amplitude associated with its cost.
\begin{equation}
    \ket{\psi} = \sum_{\vec{x}} e^{-\tau C(\vec{x})}\ket{\vec{x}},
\end{equation}
being $C(\vec{x})$ the cost associated with the combination $\vec{x}$ and $\tau$ the imaginary time evolution constant. In this way, the amplitude associated with each combination depends on its cost. Thus, for a sufficiently high value of $\tau$, the combination that minimizes the cost function has an exponentially higher tensor element than other configurations. The extraction process and demonstration are explained in detail in~\cite{melocoton}. The construction is based in applying layers of tensors that perform specific parts of the resolution. Fig.~\ref{fig: qudo_tn} shows the total tensor network with the layers that will be explained in the following. 

The tensor construction starts with a superposition layer, the first layer in Fig.~\ref{fig: qudo_tn}, where each row carries the information of the variable $x_j$ horizontally. Taking into account that each variable can take values between $0$ and $d-1$, the superposition tensor is defined as
\begin{equation}
    \ket{\psi_0} =
    \bigotimes_{i=0}^{n-1} \left(
    \sum_{x_i=0}^{d-1} \ket{x_i}
    \right)
=\sum_{\vec{x} \in \{0,\dots,d-1\}^n}\ket{\vec{x}}
\end{equation}

For example, for two variables of dimension three, the tensor is $\ket{\psi_0}=\ket{00}+\ket{01}+\ket{02}+\ket{10}+\ket{11}+\ket{12}+\ket{20}+\ket{21}+\ket{22}$. In the general case, the variable $i$-th can take values between $0$ and $d_i-1$, simply by changing the corresponding dimension of each of the tensors. However, to simplify the explanations, the rest of the paper deals only with the case where all the variables have the same range of values.

Next, the second layer in Fig.~\ref{fig: qudo_tn} computes the diagonal terms $Q_{ii}$. That is, self-interactions of the type $Q_{ii}x_ix_i$, resulting in the state
\begin{equation}
    \ket{\psi_1} = \bigotimes_{i=0}^{n-1}\left( \sum_{x_i=0}^{d-1}e^{-\tau \left(Q_{ii}x_{i}^2+D_i x_i\right)}\ket{x_i}\right).
\end{equation}

To implement the interactions between the cross variables $Q_{ij}x_ix_j$, a stair structure is proposed, in which each interaction is given by the tensors $S^{ij}$ of Fig. \ref{fig: qudo_tn}. The $C$ tensors pass the information of the variable in its row downward. Thus, the tensor $S^{ij}$ computes the interaction between the variable $x_i$ (horizontally) and $x_j$ (vertically), passing downward the value of $x_j$ and horizontally the value of $x_i$. The resultant state of applying all those layers is
\begin{equation}
     \ket{\psi_2} = e^{-\tau C(\vec{x})}\ket{\vec{x}},
\end{equation}

\begin{equation}
    e^{-\tau C(\vec{x})} = \prod_{i=0}^{n-1} \left(\phi_{ii}(x_i)\prod_{j=i+1}^{n-1} \phi_{ij}(x_i,x_j)\right),
\end{equation}

\begin{equation}
    \phi_{ii}(x_i) = e^{-\tau (Q_{ii}x_i^2+D_i x_i)},
\end{equation}
\begin{equation}
    \phi_{ij}(x_i,x_j) = e^{-\tau Q_{ij}x_ix_j}.
\end{equation}
% \begin{equation}
%     e^{-\tau C(\vec{x})} = \prod_{i=0}^{n-1} \left(e^{-\tau (Q_{ii}x_i^2+D_i x_i)}\prod_{j=i+1}^{n-1} e^{-\tau Q_{ij}x_ix_j}\right)
% \end{equation}
Each term of the product $\phi_{ij}(x_i,x_j)$ is produced by the action of the corresponding $S^{ij}$ tensor, and each term of the product $\phi_{ii}(x_i)$ is implemented by the action of its corresponding $S^{ii}$ tensor. Appendix~\ref{appendix:section-1} presents a precise definition of each tensor.

Finally, in order to determine the value of the variable $x_i$, superposition nodes are imposed on all rows but the row associated with the variable to determine. The tensor network of Fig. \ref{fig: qudo_tn} represents the equation for obtaining the first variable value. This leads to the tensor whose vector representation is
\begin{equation}
\ket{P^{0}}=\sum_{j=0}^{d-1}\left(\sum_{\vec{x}|x_0=j}e^{-\tau C(\vec{x})}\right)\ket{j}.
\end{equation}
This enables the efficient computation of a marginal `probability' of measuring each value for that variable, avoiding the storage of all possible configurations. For a sufficiently large value of $\tau$, it is possible to ensure that the value of $x_0$ is the component of $\ket{P^0}$ with a higher amplitude. Each variable $x_i$ of the solution vector has an equivalent tensor network that generates its corresponding $\ket{P^i}$ leaving free only the index of the $i$-th variable, tracing over the other indices. This means that the solution can be obtained with the iterative process proposed in \cite{melocoton}, consisting of determining each variable, eliminating in its iteration the tensors of those already determined, and introducing their value information into the remaining tensor network. These tensor networks are the exact equation that solves the QUDO problem. 

\subsubsection{Calculation and computational complexity analysis}\label{sssec:gen complexity}

\begin{figure}
    \centering
\includegraphics[width=\linewidth]{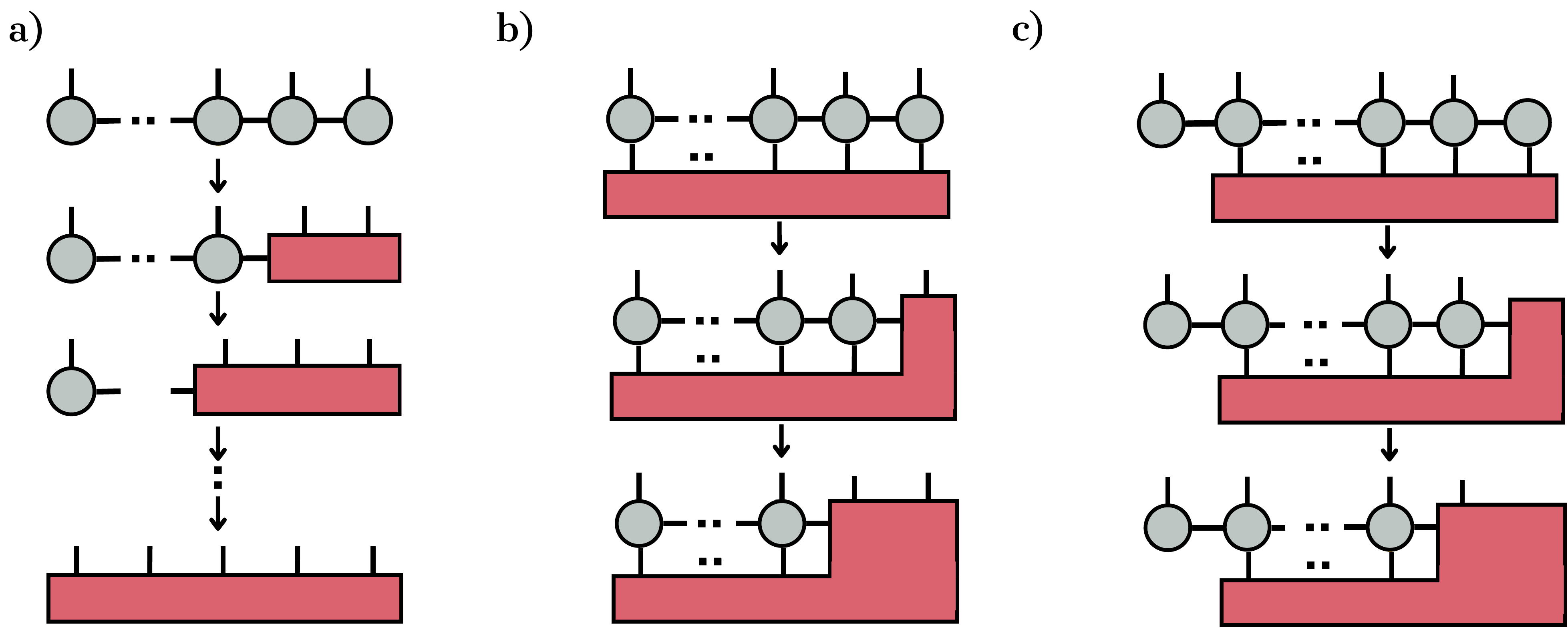}
    \caption{a) Contraction scheme of the last row MPS. b) Contraction scheme between the last row tensor and the MPO of the layer above. c) Contraction scheme between a row tensor and the MPO of the layer above in the k-neighbours case.}
    \label{fig: qudo_complexity}
\end{figure}

To compute the solution values, the tensor network must be contracted. To contract this tensor network, the contraction scheme consists in contracting the last row of the network from right to left, as shown in Fig.~\ref{fig: qudo_complexity} a, and then contracting it with the previous row, also from right to left as shown in Fig.~\ref{fig: qudo_complexity} b, following a similar approach to~\cite{TSP_TN}. To obtain the best performance, the contraction must take into account the sparsity of the tensors. Each $S$ tensor with more than two indices has only $\mathcal{O}(d^2)$ non-zero elements, and $C$ tensors have only $d$ non-zero elements. This is due to the value transmission constraints. The income value in a tensor must be the same in the outcome of the same direction. The contraction of the one- and two-indices tensors is less costly and is performed at the start, so they are neglected in the analysis.

To determine the first variable value, it is necessary to contract the entire tensor network. The contraction of the last tensor of the last row requires $\mathcal{O}(d^3)$ operations in the sparse format, versus $\mathcal{O}(d^4)$ in the full dense format. Contracting the $n-2$ last tensors of the layer requires
\begin{equation}
    \mathcal{O}\left(\sum_{m=1}^{n-3} d^2 d^m\right)=\mathcal{O}\left(d^{n-1}\right).
\end{equation}
In the full dense case, it would be
\begin{equation}
    \mathcal{O}\left(\sum_{m=1}^{n-3} d^3 d^m\right)=\mathcal{O}\left(d^{n}\right).
\end{equation}
The first tensor of the row is contracted in $\mathcal{O}(d^n)$ in both formats, so this is the total complexity of this step. The contraction of the last row with the previous one is more complex. The following considers the contraction of the $m$-th row, now a large $m$-tensor, with the previous row, with $m$ tensors. In the contraction scheme, the large tensor is contracted with the previous row from right to left, to reduce the number of free indices.

First, the last tensor of the row, which has $\mathcal{O}(d)$ non-zero elements, is contracted with the large tensor in $\mathcal{O}(d^m)$. Each contraction of the large tensor with a tensor of the previous row is performed again in $\mathcal{O}(d^m)$ operations, due to the contraction of two indices of the large tensor with the row tensor of $\mathcal{O}(d^2)$ non-zero elements. The total complexity of contracting the large tensor with a complete row is $\mathcal{O}(md^m)$ taking advantage of the sparsity. If it is contracted in a full dense format, it grows to $\mathcal{O}(d^{m+2})$. Adding the operations for all the $n$ layers, the contraction of the entire tensor network requires $\mathcal{O}(nd^{n-1})$ operations in sparse format and $\mathcal{O}(nd^{n+1})$ in full dense format, making the last two layers the bottleneck of the contraction. The following variables require smaller tensor networks, so the determination of the $m$-th variable, without reuse of intermediate computations, requires $\mathcal{O}((n-m)d^{n-1-m})$ operations with sparsity and $\mathcal{O}((n-m)d^{n+1-m})$ with full dense format. The total complexity of the algorithm is the same of the first variable determination due to this exponential reduction, with and without reuse of intermediate computations. The difference between both formats is negligible in complexity, but important in practical applications and for further algorithms in this work. Solving the general case exactly is feasible only for a very reduced number of variables due to the exponential complexity. The computational complexity of contracting the general QUDO problem of Fig. \ref{fig: qudo_tn} is $\mathcal{O}(nd^{n-1})$, and this is also the complexity of the whole algorithm.

The spatial complexity of the initial tensor network nodes is $\mathcal{O}\left(n^2d^2\right)$, because it is a lattice of $\mathcal{O}(n^2)$ tensors of $\mathcal{O}(d^2)$ non-zero elements. The tensor of the last row contracted has $n-1$ indices of dimension $d$, so it has $\mathcal{O}(d^{n-1})$ elements. This is the most costly tensor of contraction and can be rewritten in the following contraction steps. This means that the total spatial complexity of one determination step is $\mathcal{O}\left(d^{n-1}\right)$. If there is no reuse of intermediate computations, this is the total spatial complexity of the algorithm. However, with the reuse of intermediate computations, there is a need to store the tensor obtained after each row contraction step. The $m$-th row has $m$ indices, so its size is $\mathcal{O}(d^m)$. Then, the total size for storing all of them is $\mathcal{O}\left(\sum_{m=0}^{n-1} d^m\right)=\mathcal{O}\left(d^{n-1}\right)$, which is the total spatial complexity.

%The contraction of the last row (of $n-1$ indices) has a cost of $\mathcal{O}(d^{n})$, and contracting it with the previous one requires $\mathcal{O}((n-1)d^{n})$. The $m$-th row has $m$ tensors, its contraction with the following requires $\mathcal{O}(md^{m+1})$, and there are $n$ rows to contract, so summing all their costs results in $\mathcal{O}(nd^{n})$ in the determination of the first variable. Using reuse of intermediate computations, the determination of the following is less costly, keeping the complexity as $\mathcal{O}(nd^{n})$, so the bottleneck is the first.

\subsection{$k$-neighbors QUDO problems}\label{ssec: kneigh qudo}
The computational cost of contracting this tensor network grows exponentially with the number of variables, so it is not possible to contract the tensor network without employing some kind of approximation. Although there are several approximate techniques that could address the contraction of the tensor network, the remainder of the paper is restricted to solving the particular case of the $k$-neighbors in a lineal chain QUDO problem. In this case, this section introduces two alternative tensor network constructions to solve the problem, as illustrated in Figs. \ref{fig: qudo_tensor_vecinos} and \ref{fig: qudo_matrix_vecinos}.

\subsubsection{Tensor network deduction}\label{sssec:kneigh deduction}
This section approaches the $k$-neighbors QUDO problem. Basically, the difference from the general one is that each variable interacts only with its $k$ nearest neighbors in a linear chain. The cost function is expressed as
\begin{equation}
C(\vec{x}) = \sum_{i=0}^{n-1}\sum_{j=0}^{k}Q_{i,i+j}x_ix_{i+j} + \sum_{i=0}^{n-1}D_i x_i.
\end{equation}
This variant is computationally more approachable despite the size of the solution space being the same.

\begin{figure}
    \centering
    \includegraphics[width=0.95\linewidth]{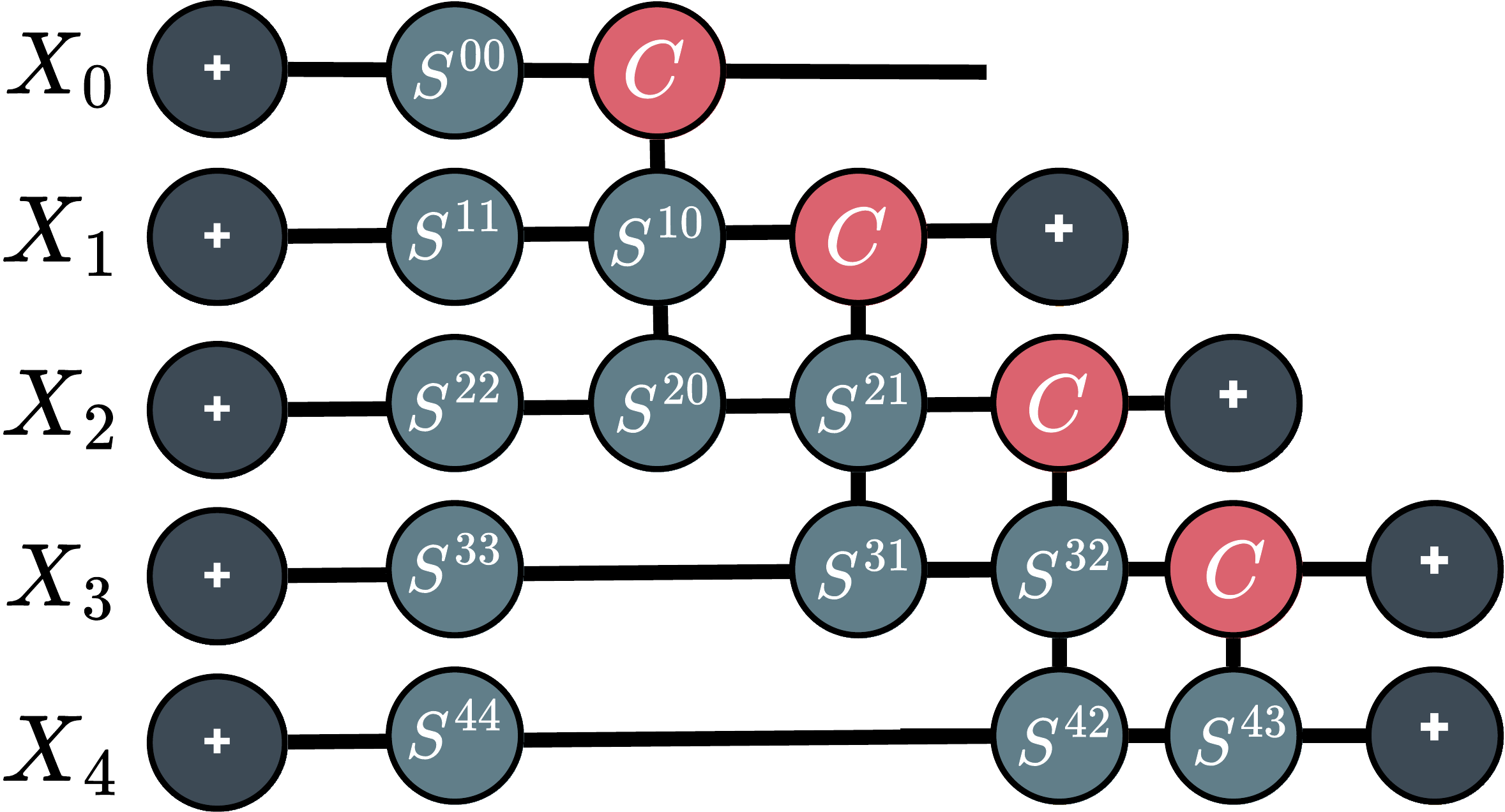}
    \caption{Tensor network for a $k$-neighbors QUDO problem with 5 variables and $k=2$. The mathematical description of each tensor is in Appendix \ref{appendix:section-1}.}
    \label{fig: qudo_tn_vecinos}
\end{figure} 

\begin{figure}
    \centering
\includegraphics[width=0.95\linewidth]{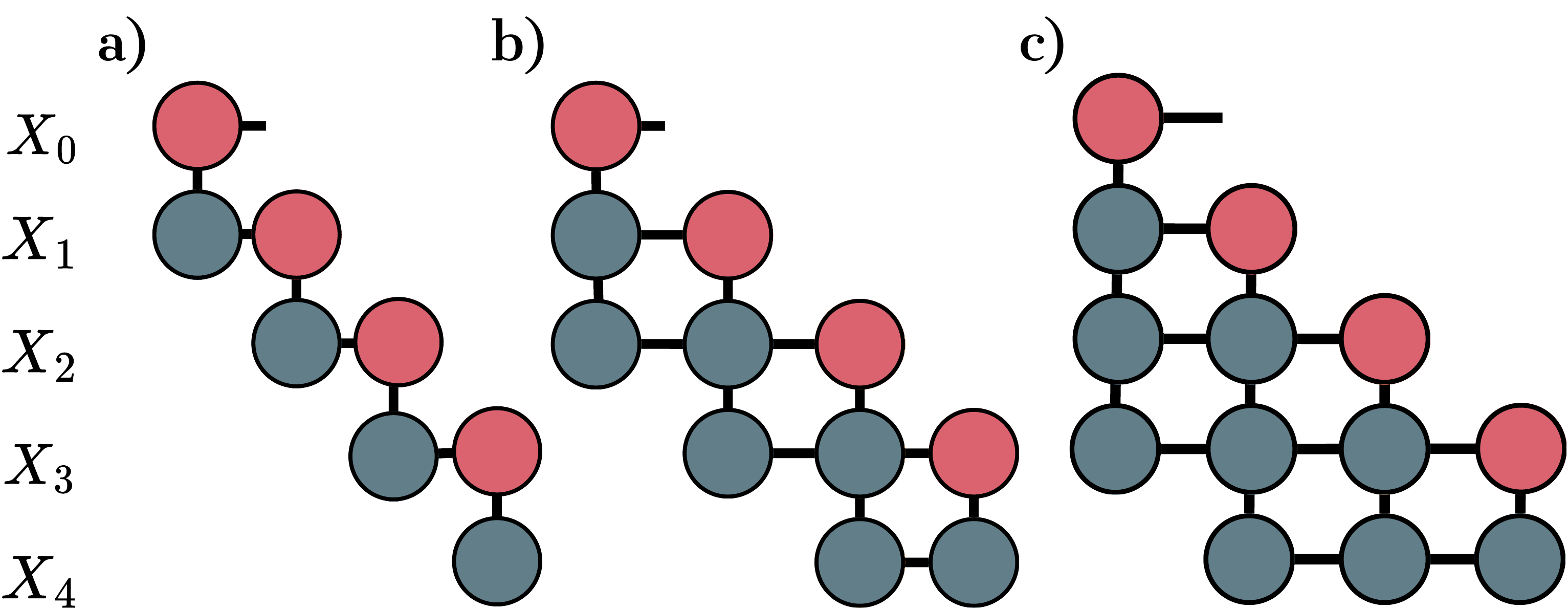}
    \caption{Schematic representation of the 5 variable $k$-neighbors QUDO tensor network for a) 1, b) 2 and c) 3-neighbors.}
    \label{fig: qudo_tensor_vecinos}
\end{figure}

The tensor network that implements this variant is similar to the network in Fig. \ref{fig: qudo_tn} by removing the interactions between the variables not connected, as shown in Fig. \ref{fig: qudo_tn_vecinos}.
It is possible to simplify this tensor network by contracting the superposition and $S^{ii}$ tensors with its corresponding $C$ and $S^{ij}$ nodes. In Fig. \ref{fig: qudo_tensor_vecinos}, the schematic tensor network associated with one, two, and three neighbors is represented. As expected, as the number of neighbors increases, the tensor network becomes more complex and therefore harder to contract. However, for problems where the number of neighbors is low enough, it is possible to contract the tensor network exactly. Following an optimal contraction scheme, it is possible to contract the entire tensor network in such a way that the computational complexity is $\mathcal{O}(knd^{k})$, as shown in Sec.~\ref{sssec:kneigh complexity}.

In addition, another implementation of the same tensor network is proposed but with a different structure. The idea is to use matrices instead of 4-order tensors because they are easier to work with and the libraries that handle them are highly optimized. Each matrix represents all the information of its row of tensors. Fig. \ref{fig: qudo_matrix_vecinos} shows how it is possible to transform from a tensor to a matrix representation.

\subsubsection{Calculation and computational complexity analysis}\label{sssec:kneigh complexity}
\begin{figure}
    \centering
    \includegraphics[width=0.85\linewidth]{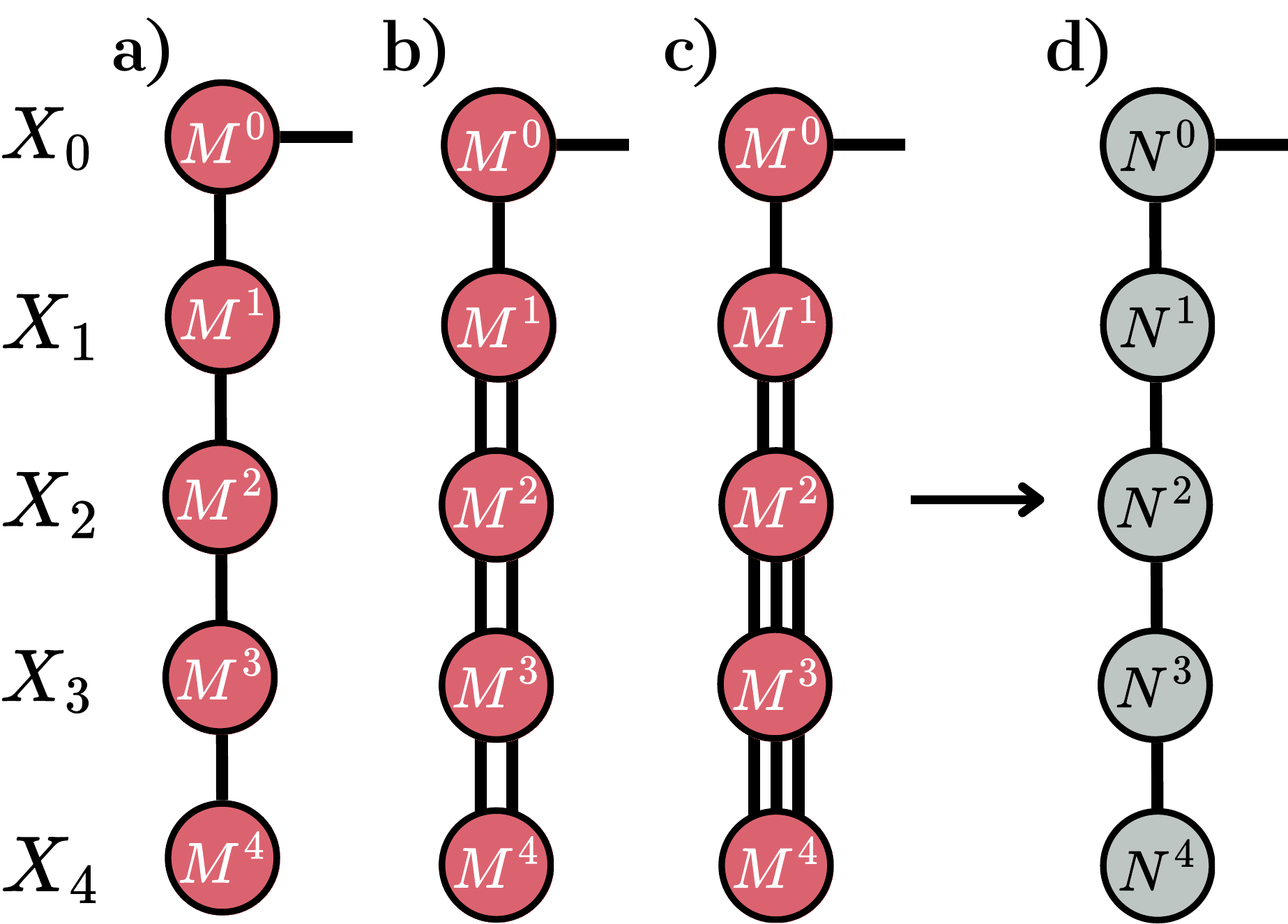}
    \caption{Schematic representation of the 5 variable $k$-neighbors QUDO tensor network for 1 a), 2 b) and 3 c) neighbors and its representation in the form of a succession of matrices d).}
    \label{fig: qudo_matrix_vecinos}
\end{figure}

It is important to keep in mind that, in the matrix method, the bond dimension depends on the number of neighbors, as $d^{k}$. For example, to translate the tensor $M^3$ in Fig. \ref{fig: qudo_matrix_vecinos} c with $6 = 3 \cdot 2$ indices into the matrix $N^3$, it is necessary to create a matrix $d^3 \times d^3$. Therefore, contracting the whole chain is equivalent to computing $n$ matrix-vector multiplications. Since determining each variable in the QUDO solution requires the contraction of a tensor chain, in principle, $n-1$ chains must be contracted. However, using the technique of taking advantage of the intermediate calculations explained in \cite{melocoton, TSP_TN}, only one contraction of the whole tensor network is needed. The same idea can be applied to the tensor method.

For the method based on matrix-vector multiplication, the computational complexity of contracting the tensor network and of the algorithm is $\mathcal{O}(nd^{2k})$.  It requires contracting $n$ times a vector with a $d^k\times d^k$ matrix in the determination of the first variable, and the following is simply a single matrix-vector multiplication of dimensions $d\times d$ per variable. This is acceptable for problems where each variable interacts only with a small number of neighbors. However, taking into account that each tensor has $k-1$ pairs of input-ouput indices with the same value, the tensors are sparse and have $\mathcal{O}(d^{k+1})$ elements, so each multiplication requires only $\mathcal{O}(d^{k+1})$ operations. With the $n$ multiplications for the complete contraction, the computational complexity per variable is $\mathcal{O}(nd^{k+1})$, being the total $\mathcal{O}(n^2d^{k+1})$ without reuse and $\mathcal{O}(nd^{k+1})$ with reuse of intermediate computations. The spatial complexity of this method is $\mathcal{O}(nd^{k+1})$, because the first tensors generated have the more non-zero elements than the following contracted vectors.

The complexity of contracting the stair structured tensor network in Fig. \ref{fig: qudo_tensor_vecinos} is more difficult to determine. In the case $k=1$, it is directly verified that the computational complexity of the algorithm using sparsity is $\mathcal{O}(nd^{2})$ and for $k=2$ is $\mathcal{O}(nd^{3})$. In cases where $k \geq 3$, the computational complexity of contracting the tensor network by rows, from bottom to top, is $\mathcal{O}(nkd^{k})$. Fig. \ref{fig: qudo_complexity} a shows how the last row is contracted. This row is made up of $k$ nodes, so the computational complexity of contracting the whole row is $\mathcal{O}(d^{k})$, resulting in a tensor like the one shown in the bottom of Fig. \ref{fig: qudo_complexity} with $k$ indices. The process of contracting this tensor with the tensor of the row above can be seen in Fig. \ref{fig: qudo_complexity} b. The contraction with the last tensor of the row requires $\mathcal{O}(d^{k})$ operations, each contraction in the row also requires $\mathcal{O}(d^{k})$ operations, and the last one requires $\mathcal{O}(d^{k+1})$ operations. The sum is $\mathcal{O}(kd^{k}+d^{k+1})$ operations, and for the $n$ rows, it grows to $\mathcal{O}(knd^{k}+nd^{k+1})$. Taking that $k$ is a parameter that can grow with the size of the problem, the complexity of the contraction of the tensor network can be reduced to $\mathcal{O}(knd^{k})$ using the sparsity of the tensors. Without sparsity, the contraction complexity is $\mathcal{O}(knd^{k+2})$. This process must be repeated for each variable, reducing the size of the tensor network. With reuse of intermediate computations, the total complexity of the algorithm is $\mathcal{O}(knd^{k})$, and without reuse, it is $\mathcal{O}(kn^2d^{k})$.

% The computational complexity of this process is $\mathcal{O}(kd^{k+2})$, because in the worst case it is necessary to contract two indices between a 4-index tensor with a $k$-tensor, and this happens $k$ times. This process has to be repeated for each row, which finally gives a computational complexity of $\mathcal{O}(knd^{k+2})$. Using reuse of intermediate computations, the following steps are cheaper, so this is the computational complexity of the algorithm.

The spatial complexity of this method implies that the $\mathcal{O}(knd^2)$ elements of the initial tensors and then the $\mathcal{O}(d^{k})$ elements of the last row tensor contracted with the last two tensors of the previous row, so the total spatial complexity is $\mathcal{O}\left(knd^2+d^{k}\right)$. Finally, reusing intermediate computations, it needs storing $n$ intermediate tensors, the spatial complexity is $\mathcal{O}(nd^{k})$.

When comparing the complexities of both methods without the use of sparsity, it can be proven that each method performs better under different numbers of neighbors and dimensions of the variables, but the tensor-based method always scales better for $k>4$. For $d=2$, both algorithms have the same scaling at $k=4$, matrix-based is better for $k<4$ and tensor-based is better for $k>4$. For $d=3$, both algorithms have the same scaling at $k=3$, matrix-based is better for $k<3$ and tensor-based is better for $k>3$. For $d>3$, matrix-based is better for $k<3$ and tensor-based is better for $k\geq 3$. However, with the use of the sparsity, the matrix-vector method is always cheaper in computational complexity. The main advantage of the tensor method is that it has lower spatial complexity, which can be interesting in certain types of instances, and the possibility of using compression methods to approximate the solution.

\subsection{Tensor QUDO extension}\label{ssec: tensor_qudo}
This algorithm can be extended to a more general problem, the Tensor Quadratic Unconstrained Discrete Optimization (T-QUDO) formulation. This problem has been approached in~\cite{melocoton} with a circular tensor network. However, it can be solved with exactly the same type of tensor network as in the QUDO case presented in Figs. \ref{fig: qudo_tn}, \ref{fig: qudo_tensor_vecinos}, and \ref{fig: qudo_matrix_vecinos}.

This time, each tensor receives the same input and passes the same output as in the QUDO case, the only change is the value of the tensor elements to implement the cost function \eqref{eq:tqudo}. Now, the tensor elements of $S^{ij}$ implement the product terms $\phi_{ij}(x_i,x_j) = e^{-\tau C_{i,j,x_i,x_j}}$ and the tensor elements of $S^{ii}$ implement the product terms $\phi_{ii}(x_i)=e^{-\tau D_{i,x_i}}$. The rest of the algorithm implementation and operation is the same.

\subsection{Exact explicit equations}\label{ssec: exact explicit}
Given these two tensor network constructions, it is possible to extract an exact and explicit equation for the problem. Given the $P^i$ tensor represented by the tensor network, the position of its largest element corresponds to the optimal value for the $i$-th variable. This means that it is possible to extract the searched value from it. However, to obtain a simple and exact equation, the value of $\tau$ must tend to infinity, which allows to avoid having to impose the results of the previous variables. $\vec{\omega}^i(\tau)$ is defined as the vector-valued function defined by the tensor network representing $P^i$, depending only on the value of $\tau$ and the definition of the problem. This implies the equation
\begin{equation}
    x_i =\arg \max_{j} \left(\lim_{\tau\rightarrow\infty} \omega^i_j(\tau)\right).
\end{equation}
However, by binarizing the variable $x_i$ into its bits $x_{i,j}$, the problem solution equation is
\begin{equation}
    x_{i,j} = \lim_{\tau\rightarrow\infty} H(\Omega_{i,j}(\tau)),
\end{equation}
being $\Omega_{i,j}(\tau)$ the tensor network obtained by contracting the tensor network of $\vec{\omega}^i$ with a $(-1,1)$ vector in the $j$-th bit index for the $i$-th variable and $H(\cdot)$ the Heaviside step function.

\section{Waterfall method}\label{sec:waterfall}

In order to extend the MeLoCoToN methodology, an alternative strategy to calculate the solution is proposed. The basic optimization in MeLoCoToN relies on taking advantage of intermediate calculations: while contracting the first tensor network to determine the value of the first variable, the intermediate tensors generated during this contraction are stored in memory. For the following variables, it is sufficient to construct a new tensor that incorporates the information of the previous known variable solutions and contracts it with the corresponding stored tensor. This approach avoids the need to perform $n$ full tensor network contractions, instead requiring only one initial contraction plus $n$ additional tensor contractions. However, this optimization entails storing $n$ intermediate tensors in memory, which can become a limiting factor for certain problem instances.
\begin{figure}
    \centering
\includegraphics[width=0.65\linewidth]{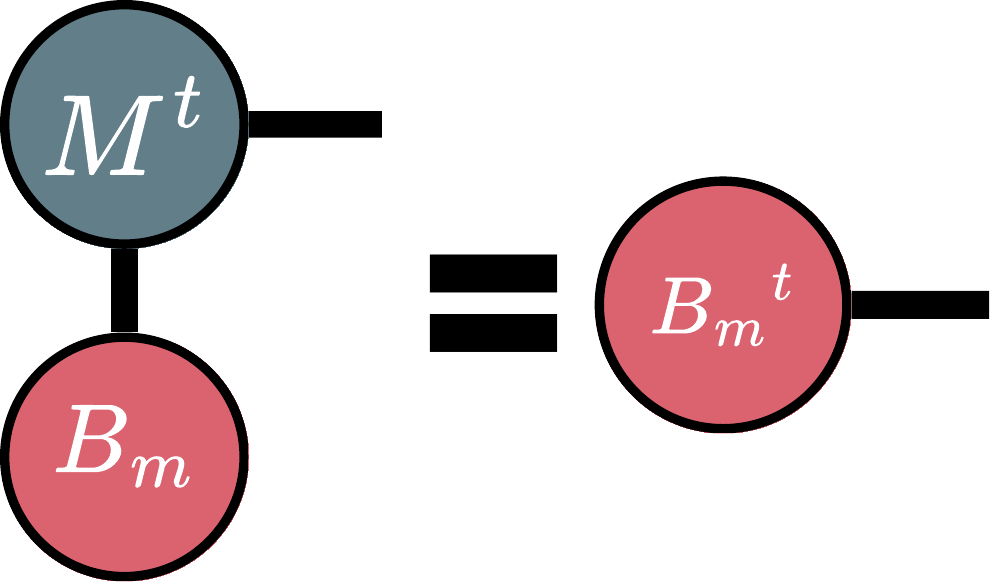}
    \caption{Contraction of the candidate tensor $M^t$ of the possible solution $t \in \{0,T\}$ and the tensor $B_m$ to obtain the tensor ${B_m}^t$.}
    \label{fig: qudo_waterfall}
\end{figure}

\begin{table*}[t]
\centering
\caption{Computational complexity for different methods and neighbor configurations. $n$ is the number of variables with integer values in $[0,d-1]$, and $k$ is the number of neighbors in the interaction.}
\label{tab:complexity}
\small
\setlength{\tabcolsep}{4pt}
\renewcommand{\arraystretch}{1.2}

\begin{tabular}{|c|c|c|c|}
\hline
\parbox[c]{5.8cm}{\centering\textbf{Method}} &
\parbox[c]{2.8cm}{\centering\textbf{Tensor method}\\\textbf{(Dense)}} &
\parbox[c]{2.8cm}{\centering\textbf{Matrix method}\\\textbf{($k$-neighbors)}} &
\parbox[c]{2.8cm}{\centering\textbf{Tensor method}\\\textbf{($k$-neighbors)}} \\
\hline

\parbox[c]{6.1cm}{\centering\textbf{Computational complexity no reuse}} &
$\mathcal{O}(n d^{n-1})$ &
$\mathcal{O}(n^2 d^{k+1})$ &
$\mathcal{O}(k n^2 d^{k})$ \\
\hline

\parbox[c]{5.8cm}{\centering\textbf{Computational complexity reuse}} &
$\mathcal{O}(n d^{n-1})$ &
$\mathcal{O}(n d^{k+1})$ &
$\mathcal{O}(k n d^{k})$ \\
\hline

\parbox[c]{6.1cm}{\centering\textbf{Computational complexity waterfall}} &
$\mathcal{O}(n d^{n})$ &
$\mathcal{O}(k n d^{k+1})$ &
$\mathcal{O}(k n d^{k+1})$ \\
\hline

\parbox[c]{5.8cm}{\centering\textbf{Spatial complexity no reuse}} &
$\mathcal{O}(d^{n-1})$ &
$\mathcal{O}(n d^{k+1})$ &
$\mathcal{O}\left(k n d^2 + d^{k}\right)$ \\
\hline

\parbox[c]{5.8cm}{\centering\textbf{Spatial complexity reuse}} &
$\mathcal{O}(d^{n-1})$ &
$\mathcal{O}(n d^{k+1})$ &
$\mathcal{O}(n d^{k})$ \\
\hline

\parbox[c]{5.8cm}{\centering\textbf{Spatial complexity waterfall}} &
$\mathcal{O}(d^{n})$ &
$\mathcal{O}\left(m d^{k+1}\right)$ &
$\mathcal{O}\left(m d^{k+1}\right)$ \\
\hline
\end{tabular}
\end{table*}

To address this limitation, a novel method for determining the solution is proposed. As in the standard MeLoCoToN approach, the tensor network is contracted from the bottom to the top. However, instead of storing the intermediate contracted tensor at each step, the method generates $T$ new vectors for each contraction level, each corresponding to a different possible configuration of the future determined variables. In other words, a separate tensor is created for every possible prior solution that the contraction process might encounter, as shown in Fig.~\ref{fig: qudo_waterfall}. For the case of $k$ neighbors, the number of tensors for the contraction is $T=d^{k}$, because that is the number of possible combinations of the values of the previous variables. The $t$-th new tensor is contracted with the current $B_m$ tensor, obtaining a new $B^t_m$ tensor, and the position $X^t_m$ of its largest element is the only stored result. So, the value $X^t_m$ is the correct result if the previous $k$ variables have the $t$-th combination of values. By evaluating all such possibilities in advance, the value of each variable for any given preceding assignment can be determined immediately and stored in a list $Y_m = (X^0_m, X^1_m,\dots, X^{T-1}_m)$. Consequently, if the value of all the previous variables (with lower $m$) is fixed to the value, the subsequent value is immediately determined without requiring further contraction. This is performed by selecting the value $x_m=X_m^t$ for the $t$-th possible combination, determined as correct for the $k$ previous variables.

A notable feature of this method is that if $Y_m$ is a constant vector of $X^t_m=X_m$ for a given variable, then the information for that variable is determined directly as $x_m = X_m$. This is because if all possible values of the previous variables yield the same result, this means that this variable value is independent of the rest of the problem, so its optimal value is trivial. Consequently, if $k=1$ and every variable depends only on the previous one, the space solution of all the next variables is automatically reduced, even under favorable conditions, and it is possible to determine the solution unambiguously. This cascading effect,  which inspires the term ``waterfall'', significantly reduces the remaining search space. Once the waterfall process is triggered, the problem effectively becomes smaller, allowing the use of larger values of the imaginary time evolution parameter $\tau$ for the remaining variables, improving the optimization and thus mitigating overflow issues.

The trade-off is an increase in computational cost: whereas the intermediate-calculation approach only requires contracting two tensors per variable after the first, the waterfall method requires $T$ contractions. For this reason, the method is best suited to problems in which $T$ is small. For example, in the $1$‑neighbor QUDO problem, the possible solutions of the previous node are $\{0, 1, \dots, d-1\}$, making the waterfall method efficient and practical in this case. However, instead of contracting the tensors to obtain the vectors, it is more simple to access the corresponding row of the tensor fixing all other indices to their variable values, which can be done in $O(1)$, and then multiplying the corresponding vector by a diagonal matrix of the extra cost associated with their $k$ neighbors interaction in $O(dk)$. In pseudocode terms
\begin{equation}
\begin{gathered}
    t \leftarrow \sum_{j=0}^{k}d^j a_j \\
    B^t_m[z] \leftarrow B_m[a_0, a_1, a_2, \dots, z] e^{-\sum_{j=0}^k Q_{m,m+j}a_{j} z},
\end{gathered}
\end{equation}
$a_j$ being the value of the $j$-th neighbor and $z$ the value of the current variable. This is repeated for each of the $T$ terms, so it has a computational complexity of $O\left(nkd^{k+1}\right)$ for the whole algorithm.

However, for $k>1$, activating the waterfall effect in the $m$-th variable for all the following variables requires that all vectors $\left\lbrace Y_m, Y_{m+1}, \dots, Y_{m+k-1}\right\rbrace$ be constant vectors. This can be improved taking into account that, if $Y_m$ is constant in the value $X_m$, then the vector $Y_{m+1}$ only requires to be constant in $X_{m+1}$ for the components where $x_m=X_m$, the vector $Y_{m+2}$ only requires to be constant in $X_{m+2}$ for the components where $x_m=X_m, x_{m+1}=X_{m+1}$, and so on. In other words, each vector requires $d$ times less components than the previous one to be equal. Nevertheless, the number of components that need to satisfy the condition remains $\mathcal{O}\left(d^k\right)$. This implies that the probability of success of the waterfall in a variable decreases as $\mathcal{O}\left(\mathcal{P}^k\right)$, $\mathcal{P}$ being the probability of obtaining a constant vector $Y_m$, making it unfeasible for large values of $k$. This means that the effectiveness of the waterfall method depends on the probability of having the same result for all new tensors generated from all possible previous configurations. The cascade probability for $k = 1$, which plays a key role in the efficiency of the method, will be empirically measured and discussed in Sec.~\ref{sec:experiments}.

% The computational complexity of this technique for both methods is the same as usual, but by multiplying each row by the corresponding $T=d^{k}$ tensors. This multiplication is in fact the selection of the value of the first $k$ indices, which requires $\mathcal{O}(d)$ operations, determining only the first a global factor, and the multiplication with a diagonal matrix of dimension $d$ in the last index, so it requires only $d$ operations. Repeating it $T$ times results in a computational complexity of $\mathcal{O}(d^{k+1})$. And for the $n$ variables, a total computational complexity of $\mathcal{O}(nd^{k+1})$, so in none of the cases it increases the computational complexity.

For spatial complexity, in each step the $k$-tensor of the contracted row is required, and the extraction of the combination returns a vector of $d$ components. From this vector, only the value $X^t_m$ is extracted. For the $T=d^{k}$ combinations, this requires a $k$-tensor, with a total of $\mathcal{O}(d^k)$ elements. In the worst case scenario, $\mathcal{O}(nd^{k+1})$ tensor elements are required, worse than the original algorithm of reuse of intermediate states. However, if the tensor is uniform at some point, all stored $Y$ tensors can be removed. If $\mathcal{P}$ is the probability of obtaining a uniform tensor, the number of tensors stored until obtaining a uniform tensor is $m\in[1,n]$, decreasing when $\mathcal{P}$ increases. Then the spatial complexity is $\mathcal{O}\left(md^{k+1}\right)$, improving the existing spatial complexity if $m<d$, without an extreme increase in computational complexity. The best-case scenario has a spatial complexity of $\mathcal{O}(d^{k+1})$, which is better than all other methods.

Table~\ref{tab:complexity} summarizes the computational and spatial complexities of the proposed methods.

\section{Experimentation}\label{sec:experiments}
This section presents a series of experiments designed to evaluate the performance of the proposed algorithms. For simplicity and analogy with the QUBO case, the term $D_ix_i$ is zero. The matrix method is implemented with Numpy \cite{numpy} and the tensor method with TensorKrowch \cite{tensorkrowch}. The sparsity of the tensors is not implemented in these tests because, in practice, the sparse method is notably slower than the full dense method. A simple version of the waterfall for $k=1$ is also implemented in Numpy, only for the matrix method. Sec.~\ref{ssec:general eval} presents the evaluation of the scaling of the runtime of the algorithms, compares them with each other, and benchmarks them against a traditional method based on a quadratic optimization implementation of OR-tools. Sec.~\ref{ssec:waterfall eval} evaluates the performance of the waterfall method to show the probability of obtaining a uniform tensor.% More tests could be performed to show the performance of the algorithms under other circumstances, but as there are many parameters involved and this paper consists of presenting the tensor network, they are left for further study.

\subsection{General algorithm evaluation}\label{ssec:general eval}
The first experiments consist of evaluating the execution time as a function of some parameters, such as the number of variables $n$, the number of neighbors $k$, and the dimension $d$. Since the algorithms depend on several parameters, only the experiments that show the most relevant behaviors have been selected.

\begin{figure*}
    \centering
    \begin{subfigure}[b]{0.48\linewidth}
        \centering
        \includegraphics[width=\linewidth]{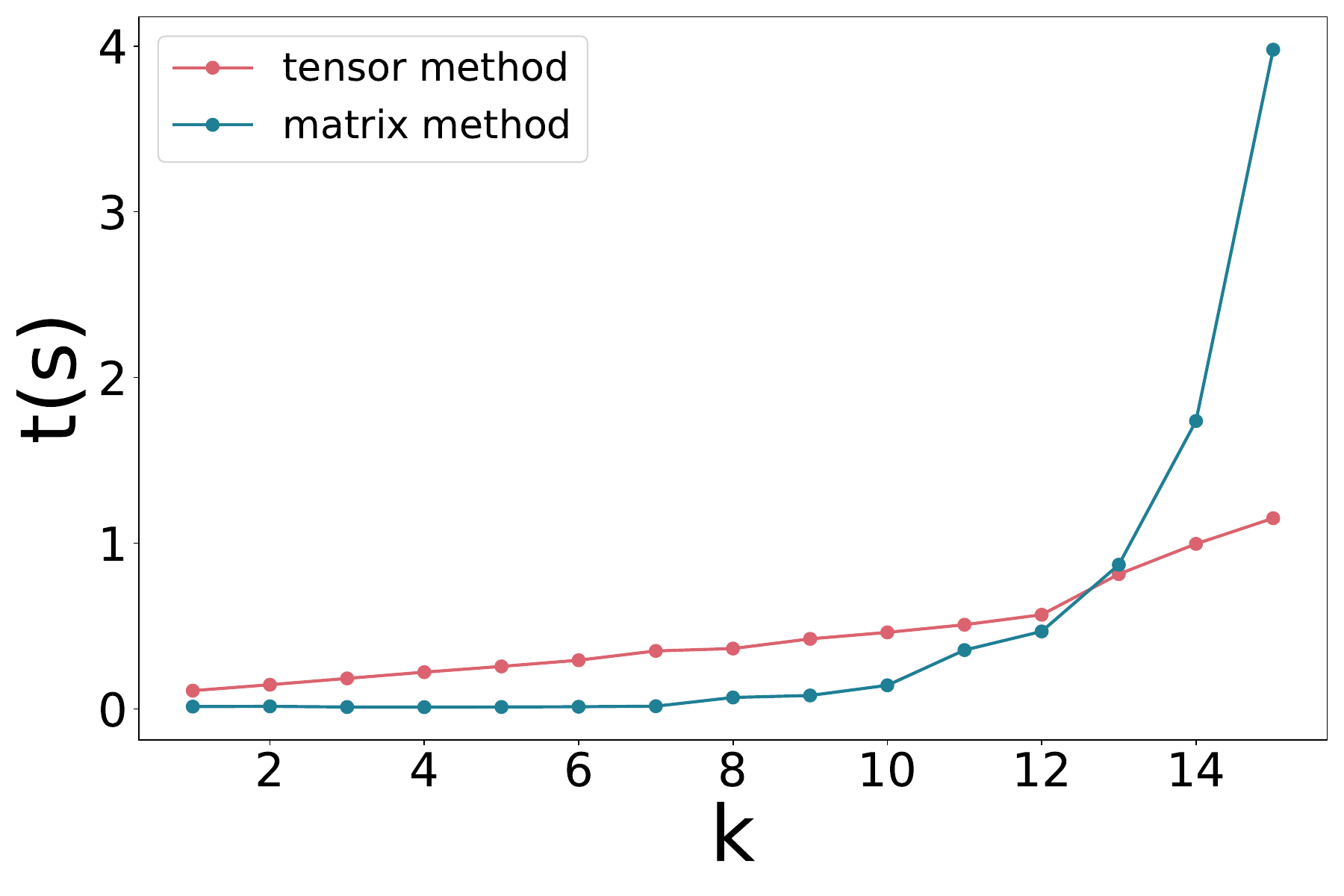}
        \caption{Execution time vs number of neighbors $k$ for the $k$-neighbors QUDO solver employing the matrix and tensor methods. $n=100$ and $d=2$.}
        \label{fig: time-vecinos}
    \end{subfigure}
    \hfill
    \begin{subfigure}[b]{0.48\linewidth}
        \centering
        \includegraphics[width=\linewidth]{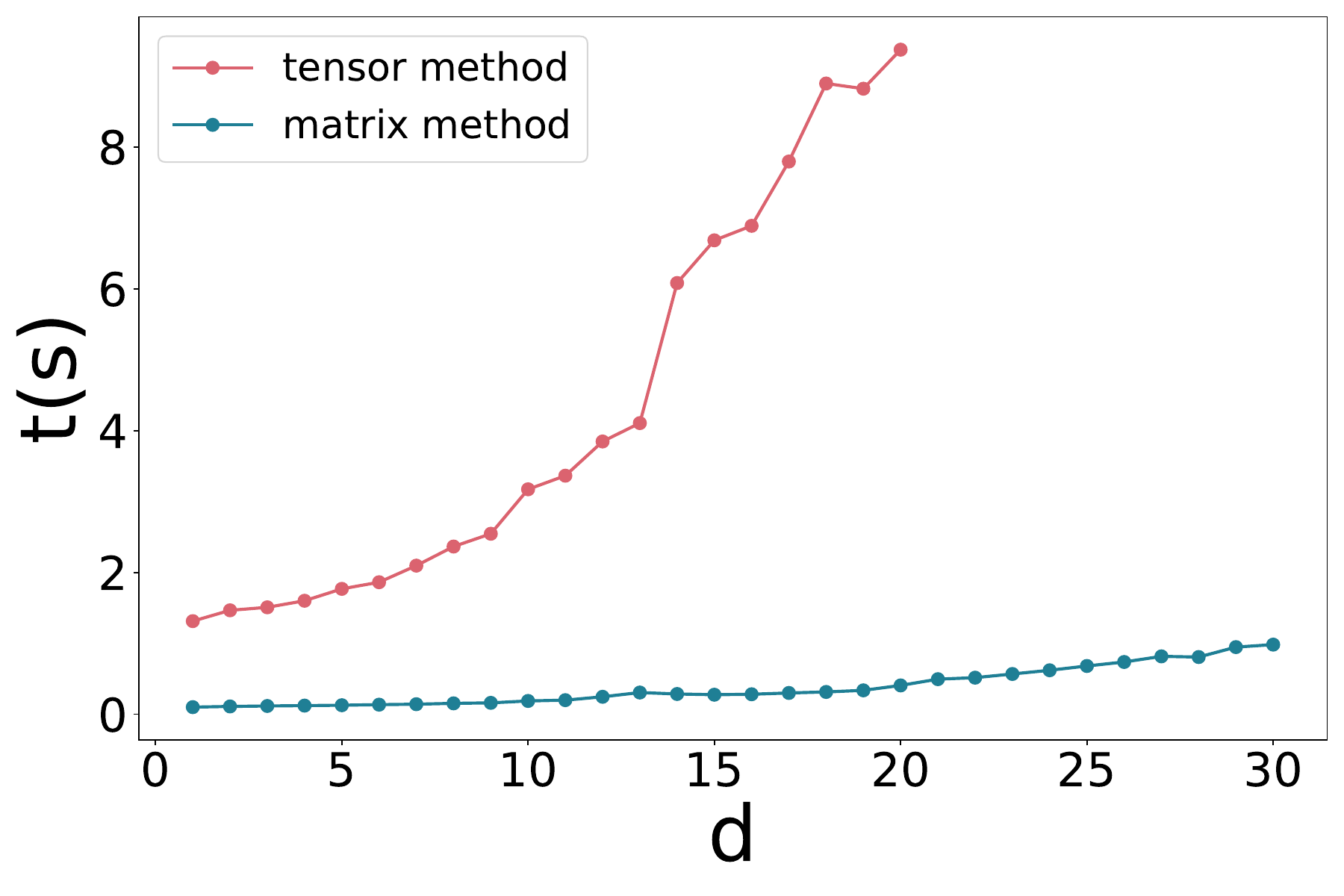}
        \caption{Execution time vs $d$ for the $k$-neighbors QUDO solver employing the matrix and tensor methods. $n=1000$ and $k=2$.}
        \label{fig: time-dits}
    \end{subfigure}
    \caption{Execution time vs $d$ and $k$ for a fixed $n$ for the QUDO solver with $\tau=400$.}
    \label{fig:time-analysis}
\end{figure*}

First, Fig. \ref{fig:time-analysis} show how the execution time $t$ of both models behaves under the dependence of $k$ and $d$. As discussed in Sec.~\ref{sssec:kneigh complexity}, in case of not using the sparsity, the tensor method scales better with $k$ and $d$. However, since the matrix implementation is more efficient despite having a worse scaling without using the sparsity, the matrix implementation is faster for sufficiently small values of $k$ and $d$. An additional comment is that increasing $k$ is computationally more expensive than increasing $d$ (recall that $n=100$ in Fig. \ref{fig: time-vecinos} and $n=1000$ in Fig. \ref{fig: time-dits}).

\begin{figure*}
    \centering
    \begin{subfigure}[b]{0.48\linewidth}
        \centering
        \includegraphics[width=\linewidth]{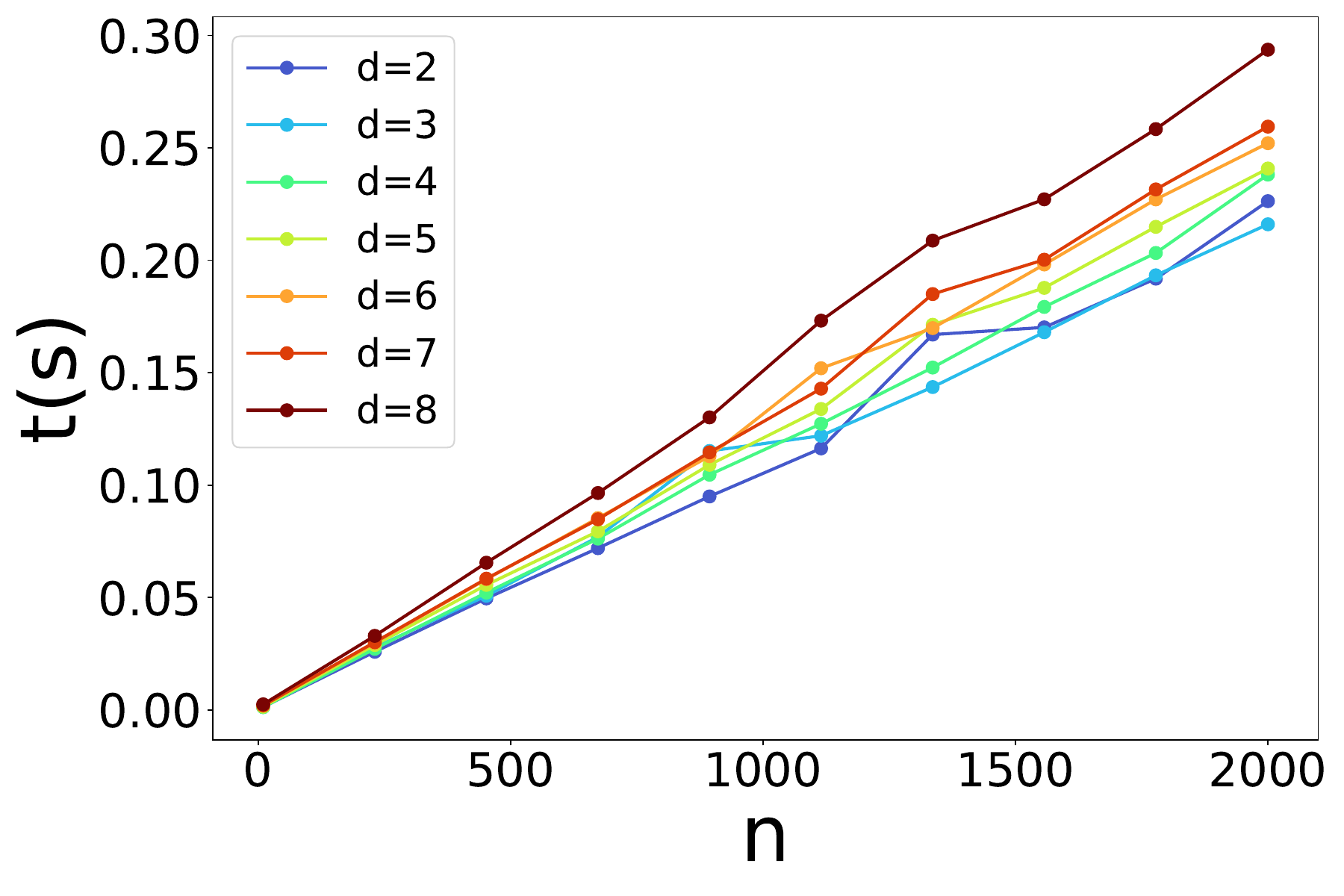}
        \caption{Matrix method comparison.}
        \label{fig: time-variables-dits-matrix}
    \end{subfigure}
    \hfill
    \begin{subfigure}[b]{0.48\linewidth}
        \centering
        \includegraphics[width=\linewidth]{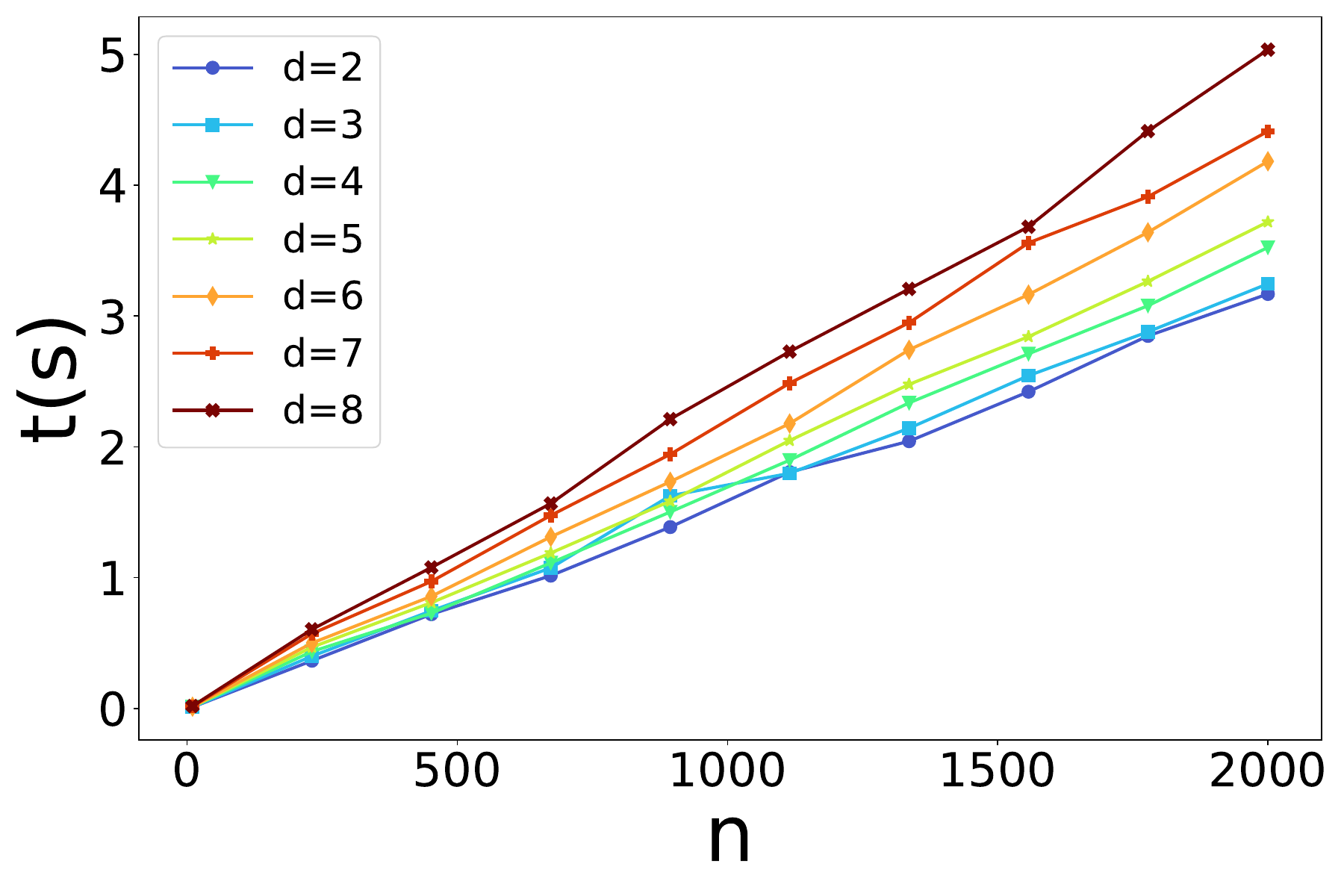}
        \caption{Tensor method comparison.}
        \label{fig: time-variables-dits-tensor}
    \end{subfigure}
    \caption{Execution time vs $n$ for the $k$-neighbors QUDO solver for different instances of $d$. $\tau=50$ and $k=2$.}
    \label{fig:time-comparison-d}
\end{figure*}

\begin{figure*}
    \centering
    \begin{subfigure}[b]{0.48\linewidth}
        \centering
        \includegraphics[width=\linewidth]{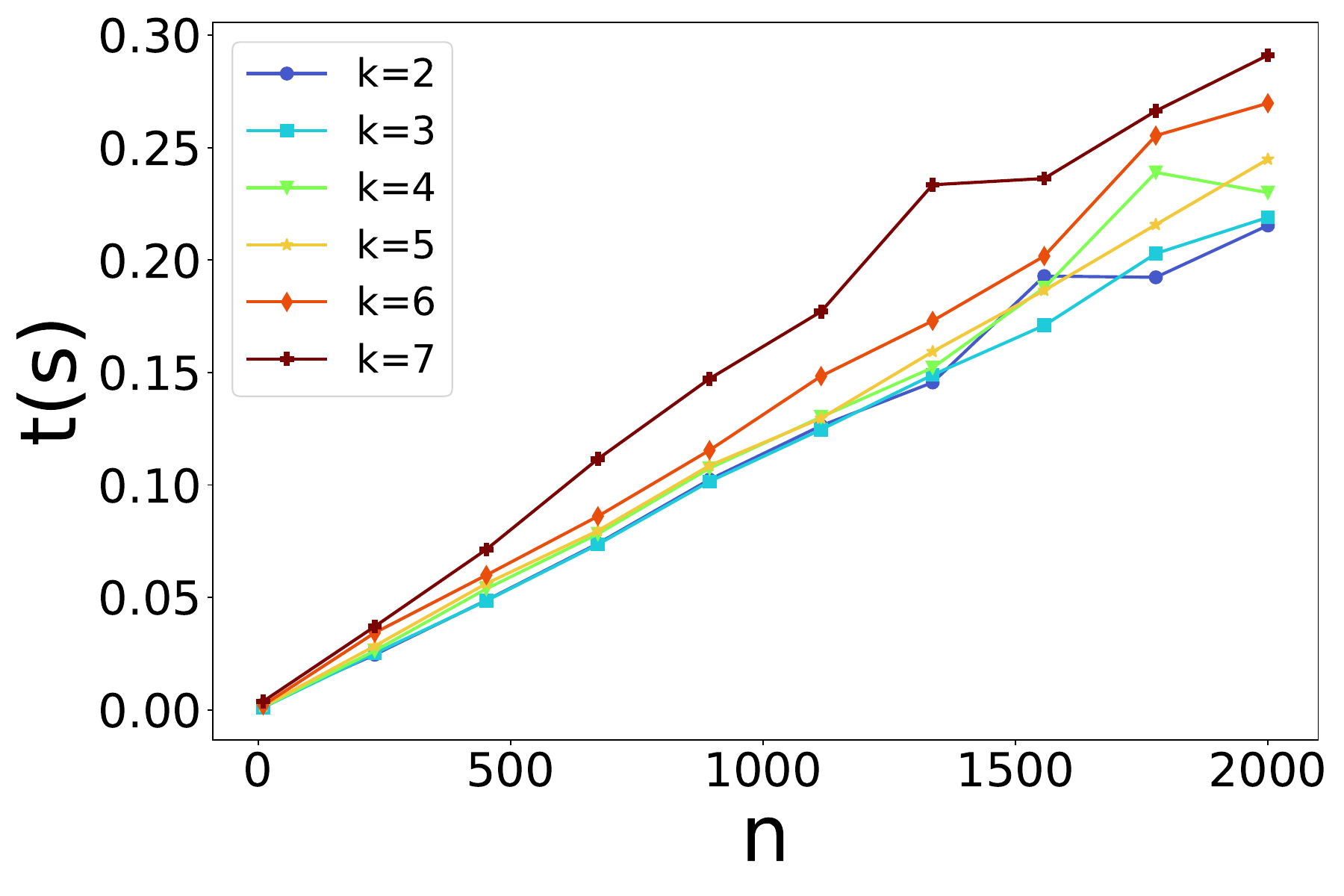}
        \caption{Matrix method comparison.}
        \label{fig: time-variables-vecinos-matrix}
    \end{subfigure}
    \hfill
    \begin{subfigure}[b]{0.48\linewidth}
        \centering
        \includegraphics[width=\linewidth]{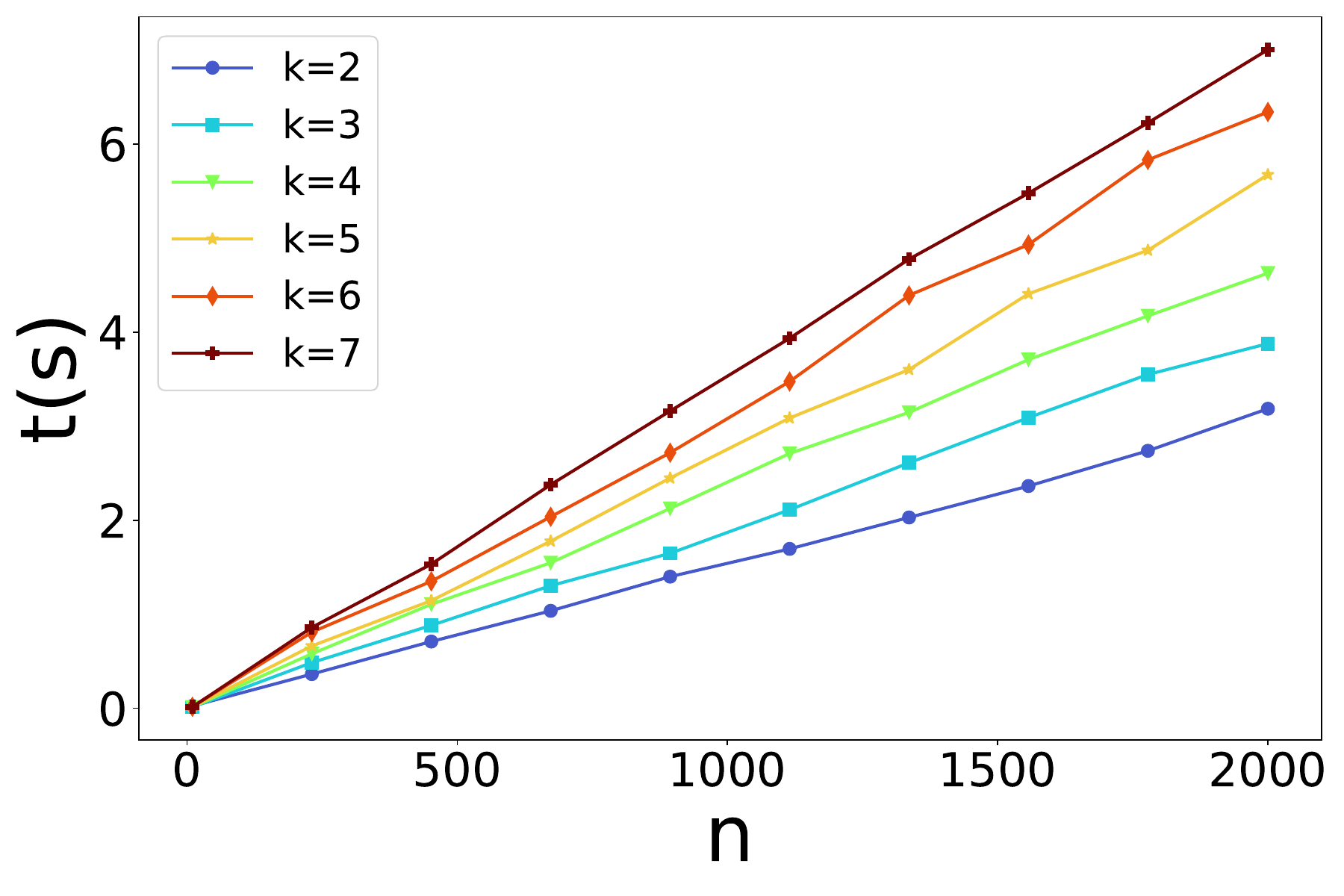}
        \caption{Tensor method comparison.}
        \label{fig: time-variables-vecinos-tensor}
    \end{subfigure}
    \caption{Execution time vs $n$ for the $k$-neighbors QUDO solver for different instances of $k$. $\tau=50$ and $d=2$.}
    \label{fig:time-comparison-k}
\end{figure*}

The following experiment evaluates the dependency between the execution time and the number of variables $n$ for both methods, shown in Figs. \ref{fig:time-comparison-d} and \ref{fig:time-comparison-k}. In particular, two different tests have been proposed, one fixing $k$ for different values of $d$, shown in Fig. \ref{fig:time-comparison-d}, and the other fixing $d$ for different values of $k$, shown in Fig. \ref{fig:time-comparison-k}.

In general, the results obtained are quite expected, and as $n$, $k$ or $d$ increases, so does the execution time. However, it is important to highlight some interesting details. Both methods show a linear behavior, which is consistent with the theoretical complexity. Additionally, for the instances used, the matrix method outperforms the tensor method in all cases. However,  this occurs only in the cases where $k$ and $d$ are sufficiently small.

\begin{figure*}
    \centering
    \begin{subfigure}[b]{0.48\linewidth}
        \centering
        \includegraphics[width=\linewidth]{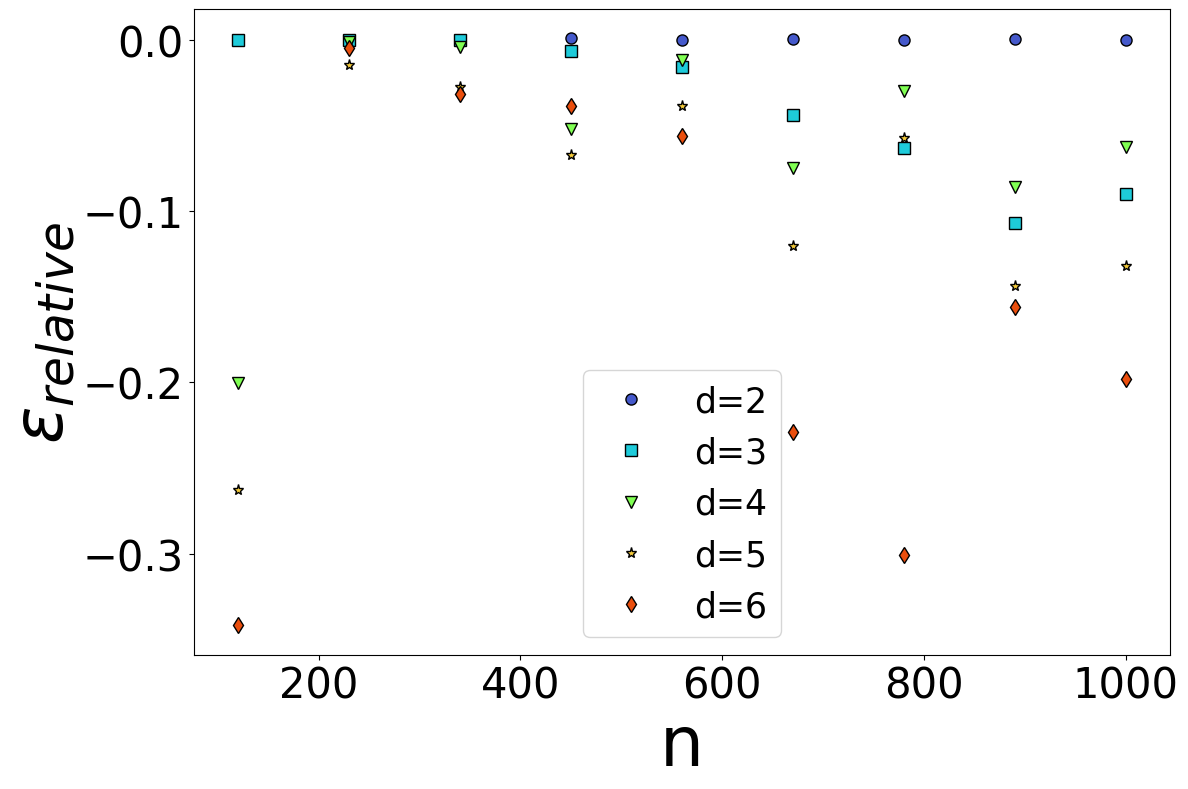}
        \caption{Relative error vs $n$ for the $k$-neighbors QUDO solver employing the matrix method for different instances of $d$. $k=2$.}
        \label{fig: rel_error-variables-dits}
    \end{subfigure}
    \hfill
    \begin{subfigure}[b]{0.48\linewidth}
        \centering
        \includegraphics[width=\linewidth]{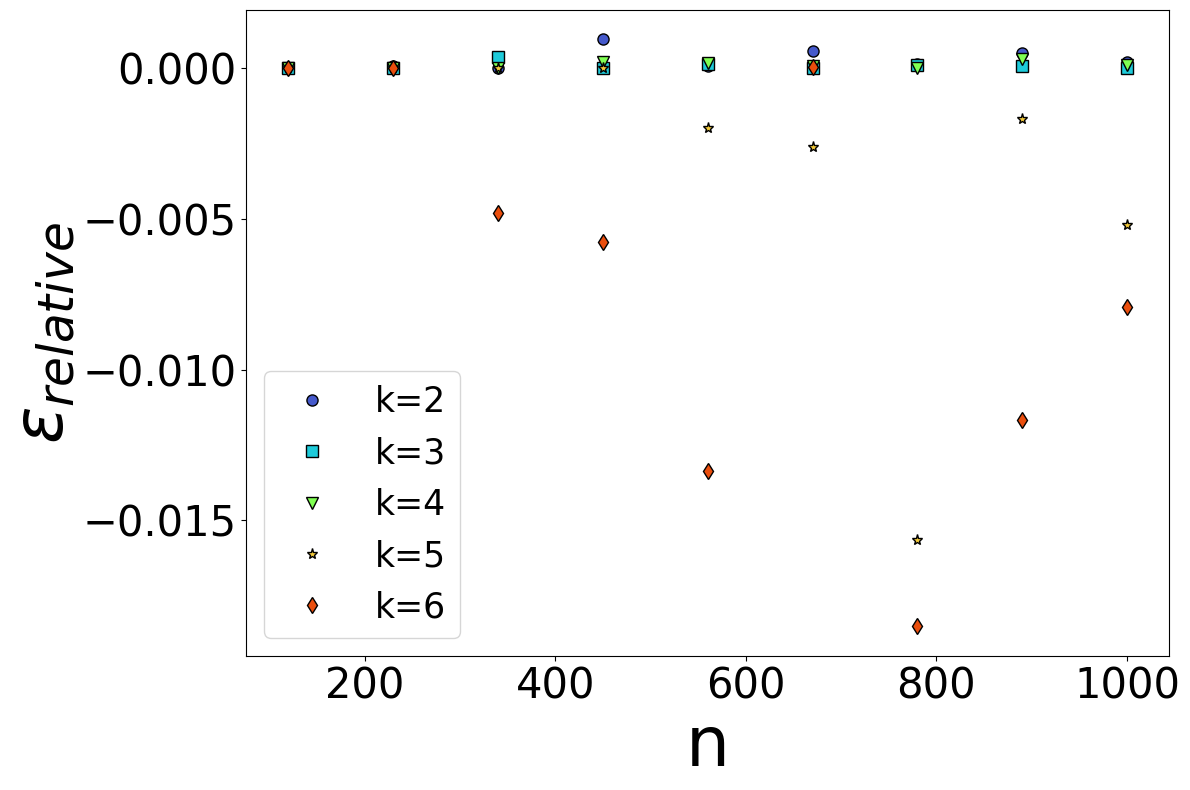}
        \caption{Relative error vs $n$ for the $k$-neighbors QUDO solver employing the matrix method for different instances of $k$. $d=2$.}
        \label{fig: rel_error-variables-vecinos}
    \end{subfigure}
    \caption{Relative error $\varepsilon_{relative}=1-\frac{C(\vec{x}_{tn})}{C(\vec{x}_{OR})}$ analysis for QUDO solver.}
    \label{fig:combined_error_analysis}
\end{figure*}

Finally, the last experiment consists of showing the accuracy of the solutions obtained by the algorithms in Figs.~\ref{fig:combined_error_analysis}. The comparison is performed against the Google OR-TOOLS solver, taking the OR-TOOLS solution as the correct solution. The search time of the OR-TOOLS solver is set to 5 seconds, because for the cases where $d>2$ the algorithm takes too long to find the most optimal solution and to make a fair comparison between methods. The selection of $\tau$ was precomputed before the algorithm was executed according to the size of the problem instance.

Since both tensor network implementations obtain exactly the same result, the following experiments are employed with the matrix method. The accuracy measure is the relative error defined as $\varepsilon_{relative}=1-\frac{C(\vec{x}_{tn})}{C(\vec{x}_{OR})}$, being $C(\vec{x}_{tn})$ the cost of the tensor network solution and $C(\vec{x}_{OR})$ the cost of the OR-TOOLS solution. Therefore, if $\varepsilon_{relative}<0$, the tensor network algorithm obtains a better solution than the OR-TOOLS. Figs.~\ref{fig:combined_error_analysis} show that the tensor network algorithm outperforms OR-TOOLS's in several instances, having at least the same performance for most of them. The experiments carried out consist of representing the dependence between $n$ and $\varepsilon_{relative}$ for some $d$ for a given $k$, shown in Fig. \ref{fig: rel_error-variables-dits}, and for some $k$ for a given $d$, shown in Fig. \ref{fig: rel_error-variables-vecinos}.

 Other optimizers have been considered, but OR-TOOLS is selected due to its fitness to the problem and availability for reproducibility. For example, dimod optimizer from D-Wave only is not general enough to tackle QUDO problems, and it provides worse solutions. Quantum computing optimizers have the same problems. GUROBI and CPLEX optimizers provide high quality solutions, but their free versions do not allow to solve instances with large number of variables. However, the increase in the quality of the solutions is due to a small $\tau$ value, so it could be solved with an adaptive method to choose it, which also explains the few instances where OR-TOOLS is better than the tensor network.

\subsection{Waterfall method evaluation}\label{ssec:waterfall eval}
This subsection presents the experimental evaluation of the waterfall method. As discussed in~\ref{sec:waterfall}, the key factor of the method is the probability of obtaining a uniform tensor and triggering the cascade effect. In particular, when a cascade event occurs and it is a 1-neighbor problem, no intermediate tensors need to be stored in memory up to that point. Moreover, the problem can be effectively `re-started' as a smaller subproblem because several variables have already been determined. This allows the selection of a larger imaginary time evolution parameter $\tau$ for the remaining variables, which can improve the accuracy of the solution.

\begin{figure}
    \centering
    \includegraphics[width=1\linewidth]{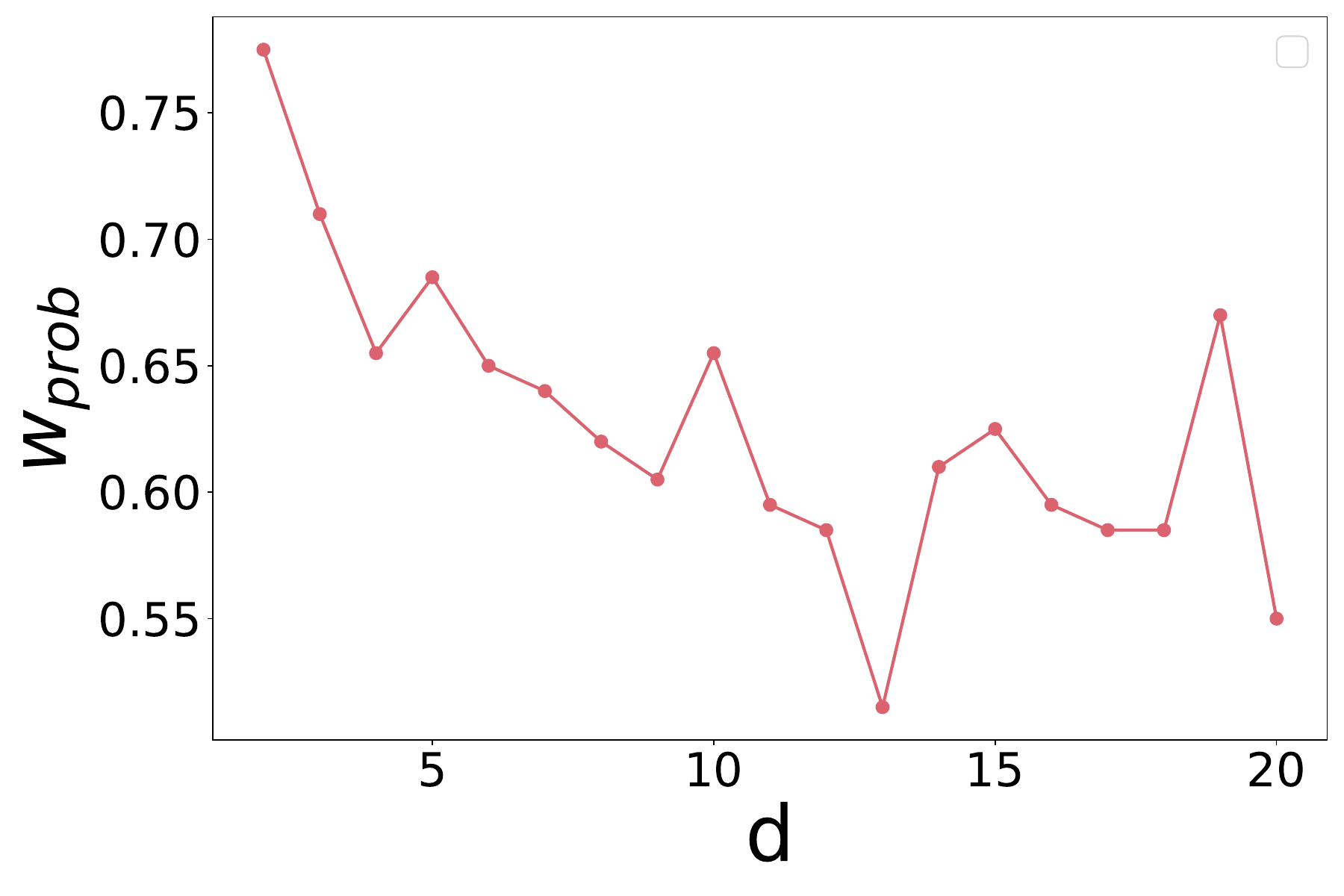}
    \caption{Waterfall averaged probability vs $d$ for the $1$-neighbor QUDO solver employing the waterfall matrix method over 50 random problem instances. Each instance is solved for 100 different values of $\tau$, and only the best solution is considered. $n=200$.}
    \label{fig: dit-watefall}
\end{figure}

To quantify this effect, the following experiment tries to empirically measure the waterfall probability for a 1-neighbor QUDO problem. The waterfall probability $w_{prob}$ is defined as the number of times the waterfall effect occurs divided by the number of nodes. Fig.~\ref{fig: dit-watefall} shows the averaged measured probability as a function of $d$ for 50 instances of random problems. The result shows a tendency for $w_{prob}$ to decrease as $d$ increases. This behavior is consistent with the fact that, as the possible dimensional solution space increases, it becomes less probable for all solutions to be the same.

\section{Conclusions}\label{sec:conclusions}
This paper presents a tensor network algorithm that solves QUBO, QUDO and Tensor-QUDO problems using the MeLoCoToN. It has been concluded that the tensor network of a dense QUDO problem has a computational complexity of $\mathcal{O}(nd^{n-1})$, which cannot be efficiently contracted. For this reason, a simpler problem with fewer density has been proposed, such as the $k$-neighbors QUDO. Two different tensor networks have been proposed, one based on tensors with up to 4 indices, the other based on matrices and vectors, and a novel methodology for reducing the required memory. The computational complexity of both implementations for a number of neighbors $k$ has been studied as $\mathcal{O}(nd^{k+1})$ (matrix method) and $\mathcal{O}(nkd^{k})$ (tensor method). In order to compare both algorithms, a series of experiments have been carried out without sparsity implementation, proving that the tensor-based method scales better as $k$ and $d$ grow. However, for problems where $k$ and $d$ take small values, the matrix-based method is faster even in this implementation. Furthermore, there is a comparison between the relative error of the results with an OR-TOOLS solver, concluding that, for most of the instances of problems, the tensor network algorithm performs equal to or better.

Future research lines may include the investigation of an efficient algorithm to determine the optimal value $\tau$ to scale the resolution and avoid the described overflow limitation. Another line may be the determination of an efficient tensor network that solves the case of $k$ interactions other than the nearest neighbors, which could be applied to several known problems. This could be approached by taking advantage of the symmetries of the problem. Also, the relation between this problem and the Ising model in physics could allow this tensor network algorithm to find the ground state of interesting physical systems. Finally, the application of compression methods in approximate representations can also be studied in order to scale the algorithm for larger or more complex problems.

\section*{Acknowledgment}
The research has been funded by the Ministry of Science and Innovation and CDTI under ECOSISTEMAS DE INNOVACIÓN project ECO-20241017 (EIFEDE) and ICECyL (Junta de Castilla y León) under project CCTT5/23/BU/0002 (QUANTUMCRIP). This project has been funded by the Spanish Ministry of Science, Innovation and Universities under the project PID2023-149511OB-I00.

% The \nocite command causes all entries in a bibliography to be printed out
% whether or not they are actually referenced in the text. This is appropriate
% for the sample file to show the different styles of references, but authors
% most likely will not want to use it.
\nocite{*}

\bibliography{apssamp}% Produces the bibliography via BibTeX.

@PREAMBLE{
 "\providecommand{\noopsort}[1]{}" 
 # "\providecommand{\singleletter}[1]{#1}%" 
}

@misc{glover2019,
      title={A Tutorial on Formulating and Using QUBO Models}, 
      author={Fred Glover and Gary Kochenberger and Yu Du},
      year={2019},
      eprint={1811.11538},
      archivePrefix={arXiv},
      primaryClass={cs.DS},
      url={https://arxiv.org/abs/1811.11538}, 
}

@book{nielsen,
  Author = {Michael A. Nielsen and Isaac L. Chuang},
  Title = {Quantum Computation and Quantum Information: 10th Anniversary Edition},
  Publisher = {Cambridge University Press},
  Year = {2011},
  ISBN = {9781107002173},
  URL = {https://www.amazon.com/Quantum-Computation-Information-10th-Anniversary/dp/1107002176?SubscriptionId=AKIAIOBINVZYXZQZ2U3A&tag=chimbori05-20&linkCode=xm2&camp=2025&creative=165953&creativeASIN=1107002176}
}

@article{Shor_1997,
   title={Polynomial-Time Algorithms for Prime Factorization and Discrete Logarithms on a Quantum Computer},
   volume={26},
   ISSN={1095-7111},
   url={http://dx.doi.org/10.1137/S0097539795293172},
   DOI={10.1137/s0097539795293172},
   number={5},
   journal={SIAM Journal on Computing},
   publisher={Society for Industrial & Applied Mathematics (SIAM)},
   author={Shor, Peter W.},
   year={1997},
   month=oct, pages={1484–1509} }

@misc{forecastingquantumcomputing,
      title={Forecasting timelines of quantum computing}, 
      author={Jaime Sevilla and C. Jess Riedel},
      year={2020},
      eprint={2009.05045},
      archivePrefix={arXiv},
      primaryClass={quant-ph},
      url={https://arxiv.org/abs/2009.05045}, 
}

@misc{herculeantask,
      title={A Herculean task: Classical simulation of quantum computers}, 
      author={Xiaosi Xu and Simon Benjamin and Jinzhao Sun and Xiao Yuan and Pan Zhang},
      year={2023},
      eprint={2302.08880},
      archivePrefix={arXiv},
      primaryClass={quant-ph},
      url={https://arxiv.org/abs/2302.08880}, 
}

@misc{tensornetworksnutshell,
      title={Tensor Networks in a Nutshell}, 
      author={Jacob Biamonte and Ville Bergholm},
      year={2017},
      eprint={1708.00006},
      archivePrefix={arXiv},
      primaryClass={quant-ph},
      url={https://arxiv.org/abs/1708.00006}, 
}

@misc{lecturesquantumtensornetworks,
      title={Lectures on Quantum Tensor Networks}, 
      author={Jacob Biamonte},
      year={2020},
      eprint={1912.10049},
      archivePrefix={arXiv},
      primaryClass={quant-ph},
      url={https://arxiv.org/abs/1912.10049}, 
}

@misc{melocoton,
      title={Explicit Solution Equation for Every Combinatorial Problem via Tensor Networks: MeLoCoToN}, 
      author={Alejandro Mata Ali},
      year={2025},
      eprint={2502.05981},
      archivePrefix={arXiv},
      primaryClass={cs.ET},
      url={https://arxiv.org/abs/2502.05981}, 
}

@article{tensorkrowch,
  title={Tensor{K}rowch: {S}mooth integration of tensor networks in machine learning},
  author={Pareja Monturiol, Jos{\'e} Ram{\'o}n and P{\'e}rez-Garc{\'i}a, David and Pozas-Kerstjens, Alejandro},
  journal={Quantum},
  volume={8},
  pages={1364},
  year={2024},
  publisher={Verein zur F{\"o}rderung des Open Access Publizierens in den Quantenwissenschaften},
  doi={10.22331/q-2024-06-11-1364},
  archivePrefix={arXiv},
  eprint={2306.08595}
}

@Article{numpy,
 title         = {Array programming with {NumPy}},
 author        = {Charles R. Harris and K. Jarrod Millman and St{\'{e}}fan J.
                 van der Walt and Ralf Gommers and Pauli Virtanen and David
                 Cournapeau and Eric Wieser and Julian Taylor and Sebastian
                 Berg and Nathaniel J. Smith and Robert Kern and Matti Picus
                 and Stephan Hoyer and Marten H. van Kerkwijk and Matthew
                 Brett and Allan Haldane and Jaime Fern{\'{a}}ndez del
                 R{\'{i}}o and Mark Wiebe and Pearu Peterson and Pierre
                 G{\'{e}}rard-Marchant and Kevin Sheppard and Tyler Reddy and
                 Warren Weckesser and Hameer Abbasi and Christoph Gohlke and
                 Travis E. Oliphant},
 year          = {2020},
 month         = sep,
 journal       = {Nature},
 volume        = {585},
 number        = {7825},
 pages         = {357--362},
 doi           = {10.1038/s41586-020-2649-2},
 publisher     = {Springer Science and Business Media {LLC}},
 url           = {https://doi.org/10.1038/s41586-020-2649-2}
}

@article{VQE,
   title={The Variational Quantum Eigensolver: A review of methods and best practices},
   volume={986},
   ISSN={0370-1573},
   url={http://dx.doi.org/10.1016/j.physrep.2022.08.003},
   DOI={10.1016/j.physrep.2022.08.003},
   journal={Physics Reports},
   publisher={Elsevier BV},
   author={Tilly, Jules and Chen, Hongxiang and Cao, Shuxiang and Picozzi, Dario and Setia, Kanav and Li, Ying and Grant, Edward and Wossnig, Leonard and Rungger, Ivan and Booth, George H. and Tennyson, Jonathan},
   year={2022},
   month=nov, pages={1–128} }

@misc{QAOA,
      title={A Quantum Approximate Optimization Algorithm}, 
      author={Edward Farhi and Jeffrey Goldstone and Sam Gutmann},
      year={2014},
      eprint={1411.4028},
      archivePrefix={arXiv},
      primaryClass={quant-ph},
      url={https://arxiv.org/abs/1411.4028}, 
}

@article{Orus_TN,
   title={A practical introduction to tensor networks: Matrix product states and projected entangled pair states},
   volume={349},
   ISSN={0003-4916},
   url={http://dx.doi.org/10.1016/j.aop.2014.06.013},
   DOI={10.1016/j.aop.2014.06.013},
   journal={Annals of Physics},
   publisher={Elsevier BV},
   author={Orús, Román},
   year={2014},
   month=oct, pages={117–158} }

@article{Orus_Many_body,
   title={Tensor networks for complex quantum systems},
   volume={1},
   ISSN={2522-5820},
   url={http://dx.doi.org/10.1038/s42254-019-0086-7},
   DOI={10.1038/s42254-019-0086-7},
   number={9},
   journal={Nature Reviews Physics},
   publisher={Springer Science and Business Media LLC},
   author={Orús, Román},
   year={2019},
   month=aug, pages={538–550} }

@misc{QUBO_tridiaonal,
      title={Polynomial-time Solver of Tridiagonal QUBO and QUDO problems with Tensor Networks}, 
      author={Alejandro Mata Ali and Iñigo Perez Delgado and Marina Ristol Roura and Aitor Moreno Fdez. de Leceta},
      year={2024},
      eprint={2309.10509},
      archivePrefix={arXiv},
      primaryClass={quant-ph},
      url={https://arxiv.org/abs/2309.10509}, 
}

@misc{TSP_TN,
      title={Traveling Salesman Problem from a Tensor Networks Perspective}, 
      author={Alejandro Mata Ali and Iñigo Perez Delgado and Aitor Moreno Fdez. de Leceta},
      year={2024},
      eprint={2311.14344},
      archivePrefix={arXiv},
      primaryClass={quant-ph},
      url={https://arxiv.org/abs/2311.14344}, 
}

@book{combinatorial_optimization_theory_papa,
author = {Papadimitriou, Christos and Steiglitz, Kenneth},
year = {1982},
month = {01},
pages = {},
title = {Combinatorial Optimization: Algorithms and Complexity},
volume = {32},
isbn = {0-13-152462-3},
journal = {IEEE Transactions on Acoustics, Speech, and Signal Processing},
doi = {10.1109/TASSP.1984.1164450}
}

@book{combinatorial_optimization_theory_garey,
author = {Garey, Michael R. and Johnson, David S.},
title = {Computers and Intractability; A Guide to the Theory of NP-Completeness},
year = {1990},
isbn = {0716710455},
publisher = {W. H. Freeman \& Co.},
address = {USA}
}

@article{quadratic_qubo,
author = {Kochenberger, Gary and Hao, Jin-Kao and Glover, Fred and Lü, Zhipeng and Wang, Haibo and Wang, Yang},
year = {2014},
month = {07},
pages = {},
title = {The unconstrained binary quadratic programming problem: A survey},
volume = {28},
journal = {Journal of Combinatorial Optimization},
doi = {10.1007/s10878-014-9734-0}
}

@misc{gonzálezbermejo2022gpsnewtspformulation,
      title={GPS: A new TSP formulation for its generalizations type QUBO}, 
      author={Saúl González-Bermejo and Guillermo Alonso-Linaje and Parfait Atchade-Adelomou},
      year={2022},
      eprint={2110.12158},
      archivePrefix={arXiv},
      primaryClass={quant-ph},
      url={https://arxiv.org/abs/2110.12158}, 
}

@inproceedings{knapsack_qubo,
author = {Bontekoe, Tariq and Phillipson, Frank and Schoot, Ward van der},
title = {Translating Constraints into QUBOs for the Quadratic Knapsack Problem},
year = {2023},
isbn = {978-3-031-36029-9},
publisher = {Springer-Verlag},
address = {Berlin, Heidelberg},
url = {https://doi.org/10.1007/978-3-031-36030-5_8},
doi = {10.1007/978-3-031-36030-5_8},
booktitle = {Computational Science – ICCS 2023: 23rd International Conference, Prague, Czech Republic, July 3–5, 2023, Proceedings, Part V},
pages = {90–107},
numpages = {18},
location = {Prague, Czech Republic}
}

@book{brach_and_bound,
  title     = {Integer and Combinatorial Optimization},
  author    = {Nemhauser, George L. and Wolsey, Laurence A.},
  year      = {1988},
  publisher = {Wiley-Interscience},
  address   = {New York},
  isbn      = {978-0471828198}
}

@incollection{Linear,
title = {8 - Linear Programming},
editor = {Jean-Michel Réveillac},
booktitle = {Optimization Tools for Logistics},
publisher = {Elsevier},
pages = {209-237},
year = {2015},
isbn = {978-1-78548-049-2},
doi = {https://doi.org/10.1016/B978-1-78548-049-2.50008-6},
url = {https://www.sciencedirect.com/science/article/pii/B9781785480492500086},
author = {Jean-Michel Réveillac}
}

@article{simmulated_annealing_kirk,
author = {S. Kirkpatrick  and C. D. Gelatt  and M. P. Vecchi },
title = {Optimization by Simulated Annealing},
journal = {Science},
volume = {220},
number = {4598},
pages = {671-680},
year = {1983},
doi = {10.1126/science.220.4598.671},
URL = {https://www.science.org/doi/abs/10.1126/science.220.4598.671},
eprint = {https://www.science.org/doi/pdf/10.1126/science.220.4598.671}}

@article{tabu_search_glover,
  title     = {Tabu Search—Part I},
  author    = {Glover, Fred},
  journal   = {ORSA Journal on Computing},
  volume    = {1},
  number    = {3},
  pages     = {190--206},
  year      = {1989},
  publisher = {INFORMS},
  doi       = {10.1287/ijoc.1.3.190}
}

@book{genetic_holland,
  title     = {Adaptation in Natural and Artificial Systems},
  author    = {Holland, John H.},
  year      = {1975},
  publisher = {University of Michigan Press},
  address   = {Ann Arbor},
  isbn      = {978-0472084609}
}

@manual{gurobi2024,
  title        = {Gurobi Optimizer Reference Manual},
  author       = {{Gurobi Optimization, LLC}},
  year         = {2025},
  url          = {https://www.gurobi.com/documentation/},
  note         = {Accessed: 2025-08-05}
}

@manual{cplex2024,
  title        = {IBM ILOG CPLEX Optimization Studio CPLEX User's Manual},
  author       = {{IBM Corporation}},
  year         = {2025},
  url          = {https://www.ibm.com/products/ilog-cplex-optimization-studio},
  note         = {Accessed: 2025-08-05}
}

@manual{ortools2024,
  title        = {OR-Tools User's Manual},
  author       = {{Google LLC}},
  year         = {2025},
  url          = {https://developers.google.com/optimization},
  note         = {Accessed: 2025-08-05}
}

@article{heuristic_comparation,
title = {A comparison between simulated annealing, genetic algorithm and tabu search methods for the unconstrained quadratic Pseudo-Boolean function},
journal = {Computers \& Industrial Engineering},
volume = {38},
number = {3},
pages = {323-340},
year = {2000},
issn = {0360-8352},
doi = {https://doi.org/10.1016/S0360-8352(00)00043-7},
url = {https://www.sciencedirect.com/science/article/pii/S0360835200000437},
author = {M. Hasan and T. AlKhamis and J. Ali}
}

@article{quantum_annealing_kadowaki,
  title     = {Quantum annealing in the transverse Ising model},
  author    = {Kadowaki, Tadashi and Nishimori, Hidetoshi},
  journal   = {Physical Review E},
  volume    = {58},
  number    = {5},
  pages     = {5355--5363},
  year      = {1998},
  publisher = {American Physical Society},
  doi       = {10.1103/PhysRevE.58.5355}
}

@book{cohen,
  title={Quantum Mechanics},
  author={Cohen-Tannoudji, Claude and Diu, Bernard and Lalo{\"e}, Franck},
  volume={1},
  year={1977},
  publisher={Wiley-Interscience}
}

@article{flex_jssp,
title = {The flexible job shop scheduling problem: A review},
journal = {European Journal of Operational Research},
volume = {314},
number = {2},
pages = {409-432},
year = {2024},
issn = {0377-2217},
doi = {https://doi.org/10.1016/j.ejor.2023.05.017},
url = {https://www.sciencedirect.com/science/article/pii/S037722172300382X},
author = {Stéphane Dauzère-Pérès and Junwen Ding and Liji Shen and Karim Tamssaouet}
}

@Article{VRP,
author={Zhang, Haifei
and Ge, Hongwei
and Yang, Jinlong
and Tong, Yubing},
title={Review of Vehicle Routing Problems: Models, Classification and Solving Algorithms},
journal={Archives of Computational Methods in Engineering},
year={2022},
month={Jan},
day={01},
volume={29},
number={1},
pages={195-221},
issn={1886-1784},
doi={10.1007/s11831-021-09574-x},
url={https://doi.org/10.1007/s11831-021-09574-x}
}

@article{Assignment,
title = {A survey of algorithms for the generalized assignment problem},
journal = {European Journal of Operational Research},
volume = {60},
number = {3},
pages = {260-272},
year = {1992},
issn = {0377-2217},
doi = {https://doi.org/10.1016/0377-2217(92)90077-M},
url = {https://www.sciencedirect.com/science/article/pii/037722179290077M},
author = {Dirk G. Cattrysse and Luk N. {Van Wassenhove}}
}

@InProceedings{quadratic,
author="Yagi, Patricia Arakawa
and Quiroz, Erik Alex Papa
and Lengua, Miguel Angel Cano",
editor="Yang, Xin-She
and Sherratt, Simon
and Dey, Nilanjan
and Joshi, Amit",
title="A Systematic Literature Review on Quadratic Programming",
booktitle="Proceedings of Seventh International Congress on Information and Communication Technology",
year="2023",
publisher="Springer Nature Singapore",
address="Singapore",
pages="739--747",
isbn="978-981-19-2397-5"
}

@Article{branch_cut,
author={Luo, Hezhi
and Chen, Sikai
and Wu, Huixian},
title={A new branch-and-cut algorithm for non-convex quadratic programming via alternative direction method and semidefinite relaxation},
journal={Numerical Algorithms},
year={2021},
month={Oct},
day={01},
volume={88},
number={2},
pages={993-1024},
issn={1572-9265},
doi={10.1007/s11075-020-01065-7},
url={https://doi.org/10.1007/s11075-020-01065-7}
}
\newpage
\appendix

\section{Tensor network definition}\label{appendix:section-1}
\begin{figure}
    \centering
    \includegraphics[width=0.8\linewidth]{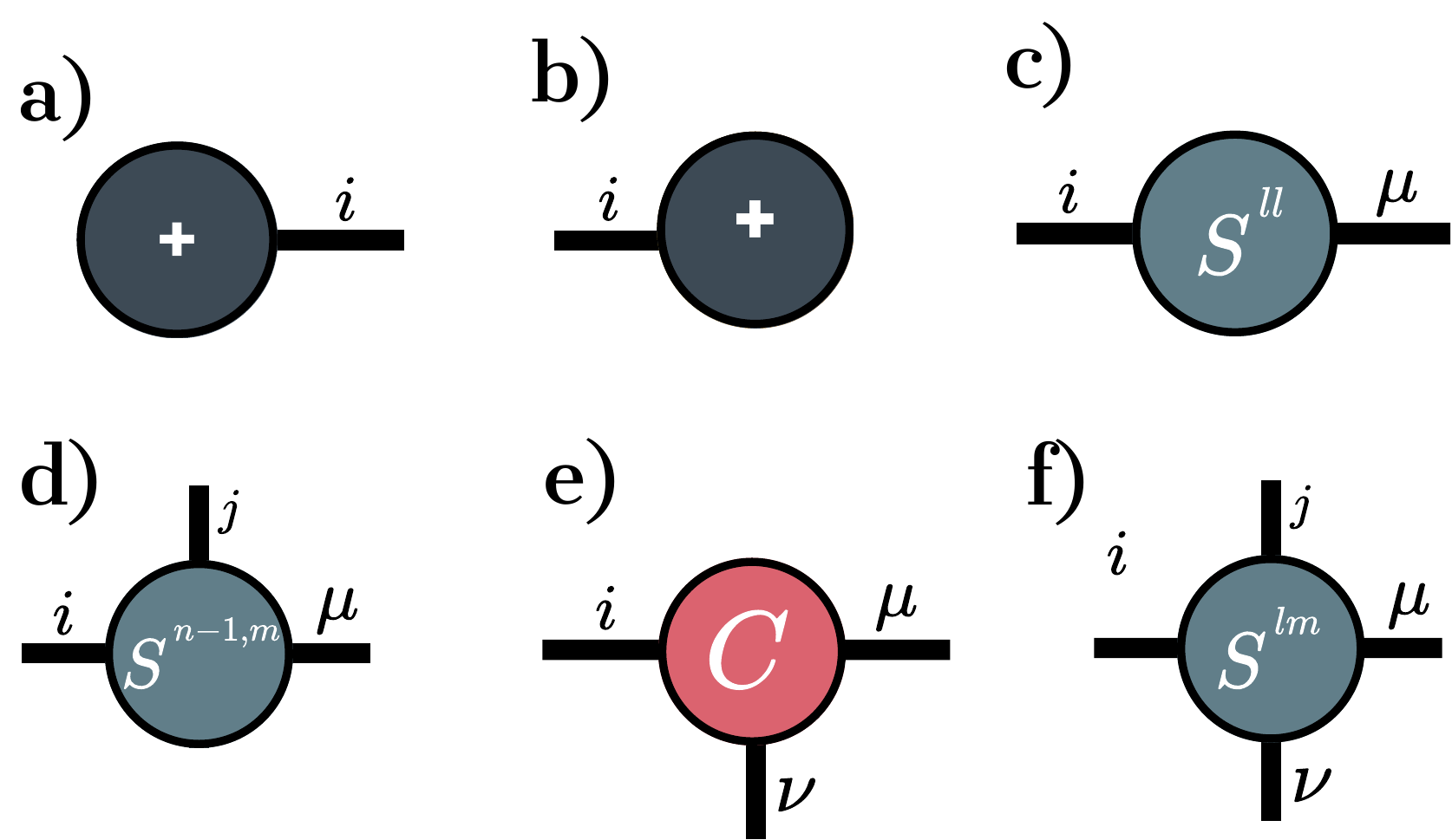}
    \caption{Tensor index naming convention.}
    \label{fig: tensor_structures}
\end{figure}

This Appendix provides the definition of the tensors used in this paper. The notation follows the conventions established in \cite{melocoton}, which should be consulted for a complete understanding. Essentially, there are five different types of tensor, shown in Fig. \ref{fig: tensor_structures}, the internal logic of each is explained below.

\paragraph{Initial superposition tensor}

$ $

The tensor $+$ corresponding to Fig. \ref{fig: tensor_structures} a, is defined as $+_{d}$
\begin{equation}
    \begin{gathered}
        +_{i} = 1.
    \end{gathered}
\end{equation}

\paragraph{Partial trace superposition tensor}

$ $

The tensor $P^+$ corresponding to Fig. \ref{fig: tensor_structures} b, is another superposition tensor as $+$. However, there is a distinction to make it clear that they are used in two different places in the tensor network. It is defined as $P^+_{d}$,
\begin{equation}
    \begin{gathered}
        P^+_{i} = 1.
    \end{gathered}
\end{equation}

\paragraph{Non-cross interaction imaginary time evolution tensor}

$ $

This tensor implements the imaginary time evolution of the QUDO interaction $Q_{ll}$ and $D_l$, shown in Fig. \ref{fig: tensor_structures} c. The tensor is defined as $S^l_{d\times d}$,  
\begin{equation}
    \begin{gathered}
        \mu = i,\\
        S^l_{i\mu} = e^{- (\tau Q_{ll} i^2+D_li)}.
    \end{gathered}
\end{equation}

\paragraph{Last row cross interaction imaginary time evolution tensor}

$ $

This tensor implements the imaginary time evolution of the QUDO interaction of the last variable $Q_{n-1,m}$, shown in Fig. \ref{fig: tensor_structures} d. The tensor is defined as $S^{n-1,m}_{d\times d \times d}$,  
\begin{equation}
    \begin{gathered}
        \mu = i \\
        S^{n-1,m}_{i\mu j} = e^{- \tau Q_{n-1,m} ij}.
    \end{gathered}
\end{equation}

\paragraph{3-index control tensor}

$ $

This tensor, shown in Fig. \ref{fig: tensor_structures} e, receives the information from $i$ and sends it without modifying it through $\mu$ and $\nu$. It is defined as $C_{d\times d\times d}$,
\begin{equation}
    \begin{gathered}
        \mu = \nu = i \\
        C_{i\mu \nu} = 1.
    \end{gathered}
\end{equation}

\paragraph{Cross interaction imaginary time evolution tensor}

$ $

This tensor implements the imaginary time evolution of the qubo interaction $Q_{lm}$ (Fig. \ref{fig: tensor_structures} f). The tensor is defined as $S^{lm}_{d\times d \times d \times d}$,  
\begin{equation}
    \begin{gathered}
        \mu = i \\
        \nu = j\\
        S^{lm}_{i\mu j \nu} = e^{- \tau Q_{lm} ij}.
    \end{gathered}
\end{equation}

\end{document}